\documentclass[prd,onecolumn,nofootinbib,aps,tightenlines,preprintnumbers,notitlepage,longbibliography]{revtex4-1}

\usepackage{amsmath}
\usepackage{amsfonts}
\usepackage{dsfont}
\usepackage{enumerate}
\usepackage{color}
\usepackage{graphicx}
\usepackage{afterpage}
\usepackage[colorlinks=true,citecolor=blue,urlcolor=blue]{hyperref}

\newcommand{\units}[1]{\ensuremath{\mathrm{#1}}}
\newcommand{\TeV}{\units{TeV}}
\newcommand{\GeV}{\units{GeV}}
\newcommand{\MeV}{\units{MeV}}
\newcommand{\keV}{\units{keV}}
\newcommand{\cm}{\units{cm}}
\newcommand{\fm}{\units{fm}}
\newcommand{\km}{\units{km}}

\newcommand{\s}{\units{s}}

\newcommand{\Msq}{\left\langle | \mathcal{M} |^2 \right\rangle}

\begin{document}

\title{First Cosmological Constraint on the\\ Effective Theory of Dark Matter--Proton Interactions}
\author{Kimberly K. Boddy}
\affiliation{
Department of Physics \& Astronomy, Johns Hopkins University, Baltimore, MD 21218, USA\\
Department of Physics \& Astronomy, University of Hawaii, Honolulu, HI 96822, USA
}
\author{Vera Gluscevic}
\affiliation{School of Natural Sciences, Institute for Advanced Study, Einstein Drive, Princeton, NJ 08540, USA\\
Department of Physics, University of Florida, Gainesville, Florida 32611, USA\\
Department of Physics, Princeton University, Princeton, NJ 08544
}

\begin{abstract}
  We obtain the first cosmological constraints on interactions between dark matter and protons within the formalism of nonrelativistic effective field theory developed for direct detection.
For each interaction operator in the effective theory, parameterized by different powers of the relative velocity of the incoming particles, we use the \textit{Planck} 2015 cosmic microwave background (CMB) temperature, polarization, and lensing anisotropy to set upper limits on the scattering cross section for all dark matter masses above 15 keV.
We find that for interactions associated with a stronger dependence on velocity, dark matter and baryons stay thermally coupled for longer, but the interaction strengths are suppressed at the low temperatures relevant for \textit{Planck} observations and are thus less constrained.
At the same time, cross sections with stronger velocity dependencies are more constrained in the limit of small dark matter mass.
In all cases, the effect of dark matter--proton scattering is most prominent on small scales in the CMB power spectra and in the matter power spectrum, and we thus expect substantial improvement over the current limits with data from ground-based CMB experiments and galaxy surveys.
\end{abstract}
\maketitle

\section{Introduction}

There are various astrophysical, cosmological, and collider-based probes engaged in an effort to detect interactions between dark matter (DM) and Standard Model particles.
The most sensitive low-energy constraints on DM interactions typically come from direct searches for rare collisions between DM particles from the local Galactic halo and atomic nuclei in low-background underground detectors~\cite{2017PhRvL.119r1301A,2017arXiv171003572F}.
Nuclear-recoil measurements from these experiments exclude large portions of the relevant parameter space, especially for DM particle masses above a few \GeV~\cite{2013arXiv1310.8327C}.
Data from a new generation of direct-detection experiments are forthcoming from a number of projects worldwide and are expected to deliver another order of magnitude in sensitivity beyond the current detection limits~\cite{2013arXiv1310.8327C}.

However, information about DM gained from direct-detection measurements is subject to several caveats.
First, the conversion of the observed nuclear-recoil rate into a limit on the DM--baryon interaction cross section relies on detailed knowledge of DM astrophysical parameters, such as the local energy density and velocity distribution.
N-body simulations show departures from the usual assumptions of a Maxwell-Boltzmann phase space distribution on small scales~\cite{2009MNRAS.395..797V,2013PhR...531....1S}, but the only way to directly measure these parameters would be through direct-detection experiments.
Second, nuclear recoil-based measurements are not sensitive to DM masses below a few GeV; sub-GeV DM would need to travel at speeds larger than the escape velocity of the Galactic halo in order to produce a detectable nuclear recoil~\cite{2013arXiv1310.8327C}.
While electronic recoil measurements can probe the parameter space of sub-GeV DM~\cite{2012PhRvL.109b1301E,2017PhRvD..96d3017E}, new technologies are necessary to address scattering with nucleons in this mass regime~\cite{Battaglieri:2017aum}.
Finally, there is a ``ceiling'' on the cross sections that underground experiments are able to probe, due to their extensive shielding: DM particles that interact too strongly would lose momentum before reaching the detector~\cite{2005PhRvD..72h3502Z,Emken:2017qmp,2017JCAP...12..004S}.

Alternative probes of low-energy DM physics can offer complementary information and may sidestep some of the aforementioned caveats of direct detection.
For example, direct-detection surface runs~\cite{Davis:2017noy}, molecular spectroscopy~\cite{Fichet:2017bng}, and balloon-borne experiments~\cite{2007PhRvD..76d2007E} have a higher a ``ceiling'' and place bounds on large interactions between DM particles and baryons.
There are also constraints that arise from the Galactic structure~\cite{2001sddm.symp..263W}, observations of galaxy clusters~\cite{2008MNRAS.384..814H,2002ApJ...580L..17N}, cosmic rays~\cite{2002PhRvD..65l3503C,2017APS..APR.J5008C}, and other astrophysical observations~\cite{1990PhRvD..41.3594S,2005A&A...438..419B,2007PhRvD..76d3523M,Kavanagh:2017cru}.
Additionally, a recent proliferation of precision cosmological measurements provides a testing ground for the very same interaction physics, but in the early Universe.
Previous studies used the cosmic microwave background (CMB) spectral distortion limit from COBE FIRAS to constrain the DM--baryon scattering for DM masses less than 100 \keV~\cite{Ali-Haimoud:2015pwa}.
References~\cite{Chen:2002yh,Sigurdson:2004zp,Dvorkin:2013cea,Gluscevic:2017ywp} used CMB temperature anisotropy measurements to constraint DM--proton interactions; Ref.~\cite{Dvorkin:2013cea} considered the limit in which DM is much heavier than the proton, and our previous work~\cite{Gluscevic:2017ywp} covered masses between a keV and a TeV, but only for velocity-independent contact interactions.

In this study, we expand on previous work to search for cosmological evidence of any nonrelativistic DM--proton effective interaction (including velocity-dependent interactions) for particle masses above 15 \keV.\footnote{Warm DM limits exclude masses below a few \keV~\cite{Viel:2013fqw}, and power spectra computations become progressively more difficult to perform at high accuracy for scattering with stronger velocity dependencies and for DM with lower masses. For these reasons, we focus on masses greater than 15 \keV.}
For this purpose, we use the latest temperature, polarization, and lensing anisotropy measurements from the \textit{Planck} 2015 data release~\cite{2016A&A...594A...1P,2016A&A...594A..11P}.
We adopt the effective theory formalism, originally developed for DM direct detection, which renders our results directly comparable to those from laboratory experiments and provides a framework to systematically investigate all possible low-energy DM interactions with protons.
We include DM scattering with helium nuclei, accounting for the nuclear responses triggered by different interaction operators.
We find no evidence for either velocity-independent or velocity-dependent DM scattering and thus present state-of-the-art cosmological constraints on DM--proton interactions; the key results are summarized in Figure~\ref{fig:exclusions}.

In Sec.~\ref{sec:EFT}, we review the nonrelativistic effective theory of DM interactions with baryons.
In Sec.~\ref{sec:cosmo}, we embed this formalism into the Boltzmann equations that describe the evolution of cosmological perturbations, allowing for the presence of DM--proton scattering in the early Universe.
In Sec.~\ref{sec:planck}, we describe the data and present our results.
In Sec.~\ref{sec:conclusions}, we discuss our results and future avenues of investigation.

\section{Dark Matter Effective Field Theory}
\label{sec:EFT}

The effective field theory (EFT) for DM interactions with nucleons enables a systematic description of processes relevant for probes of low-energy DM physics~\cite{Fan:2010gt,Fitzpatrick:2012ix,Anand:2013yka}.
In Sec.~\ref{sec:formalism}, we summarize the general EFT approach following Refs.~\cite{Fitzpatrick:2012ix,Anand:2013yka} and express the scattering amplitude for each interaction operator in a form that is useful in a cosmological setting.
In Sec.~\ref{sec:cross-sections}, we derive the associated momentum-transfer cross sections, relevant for investigating the effect of DM scattering on cosmological observables.

\subsection{Formalism}
\label{sec:formalism}

We begin by considering the nonrelativistic elastic scattering between a DM particle and a nucleon.
The complete set of Hermitian observables that describes the scattering process is as follows~\cite{Fitzpatrick:2012ix}: $i\vec{q}/m_N$ is the momentum transfer per nucleon mass; $\vec{v}^\perp$$\equiv$$\vec{v} + \vec{q}/(2\mu_{\chi N})$ is the relative velocity in a direction perpendicular to the momentum exchange (and $\mu_{\chi N}$ is the reduced mass of the DM--nucleon system); and $\vec{S}_\chi$ and $\vec{S}_N$ are the DM spin and the nucleon spin, respectively.
The momentum transfer incorporates the angular information of the scattering process via $|\vec{q}|^2$=$2\mu_{\chi N}^2 v^2 (1-\cos\theta)$, where $v$$\equiv$$|\vec{v}|$ and $\theta$ is the scattering angle in the center-of-mass frame.
The maximum possible momentum transfer is $|\vec{q}|_\textrm{max}$=$2\mu_{\chi N} v$.
Working to second order in momenta and velocities, various combinations of these four quantities give rise to the following 14 operators,\footnote{Following Ref.~\cite{Anand:2013yka}, we fix $c_2^\tau$=$0$ in order to omit $\mathcal{O}_2$=$|\vec{v}^\perp|^2$, as it does not arise at leading order from the nonrelativistic reduction of a relativistic operator; conversely, we keep $\mathcal{O}_{15}$, which is third order in momenta and velocities, because it can arise at leading order.} derived in Ref.~\cite{Anand:2013yka}:
\begin{align}
  \mathcal{O}_1    &= 1_\chi 1_N &
  \mathcal{O}_9    &= \vec{S}_\chi \cdot \left(\vec{S}_N \times \frac{i\vec{q}}{m_N}\right) \nonumber\\
  \mathcal{O}_3    &= \vec{S}_N \cdot \left(\frac{i\vec{q}}{m_N} \times \vec{v}^\perp \right) &
  \mathcal{O}_{10} &= \vec{S}_N \cdot \frac{i\vec{q}}{m_N} \nonumber\\
  \mathcal{O}_4    &= \vec{S}_\chi \cdot \vec{S}_N &
  \mathcal{O}_{11} &= \vec{S}_\chi \cdot \frac{i\vec{q}}{m_N} \nonumber\\
  \mathcal{O}_5    &= \vec{S}_\chi \cdot \left(\frac{i\vec{q}}{m_N} \times \vec{v}^\perp \right) &
  \mathcal{O}_{12} &= \vec{S}_\chi \cdot \left(\vec{S}_N \times \vec{v}^\perp\right) \nonumber\\
  \mathcal{O}_6    &= -\left(\vec{S}_\chi \cdot \frac{i\vec{q}}{m_N}\right) \left(\vec{S}_N \cdot \frac{i\vec{q}}{m_N}\right) &
  \mathcal{O}_{13} &= \left(\vec{S}_\chi \cdot \vec{v}^\perp\right) \left(\vec{S}_N \cdot \frac{i\vec{q}}{m_N}\right) \nonumber\\
  \mathcal{O}_7    &= \vec{S}_N \cdot \vec{v}^\perp &
  \mathcal{O}_{14} &= \left(\vec{S}_\chi \cdot \frac{i\vec{q}}{m_N}\right) \left(\vec{S}_N \cdot \vec{v}^\perp\right) \nonumber\\
  \mathcal{O}_8    &= \vec{S}_\chi \cdot \vec{v}^\perp &
  \mathcal{O}_{15} &= \left(\vec{S}_\chi \cdot \frac{i\vec{q}}{m_N}\right) \left[\left(\vec{S}_N \times \vec{v}^\perp\right) \cdot \frac{i\vec{q}}{m_N} \right] \ .
  \label{eq:operators}
\end{align}
Since DM may interact with both protons and neutrons, it is convenient to work in an isospin basis, in which the interaction Hamiltonian has the form
\begin{equation}
  \mathcal{H} = \sum_{\tau=0}^1 \sum_{i=1}^{15} c_i^\tau \mathcal{O}_i t^\tau \ ,
\end{equation}
where $i$ labels the interaction and $\tau$ labels isospin.
The isospin operators are $t^0$$\equiv$$1$ and $t^1$$\equiv$$\tau_3$, and the isospin coupling coefficients are $c_i^0$ and $c_i^1$.
The coupling coefficients to protons $c_i^{(p)}$=$(c_i^0 + c_i^1)/2$ and neutrons $c_i^{(n)}$=$(c_i^0 - c_i^1)/2$ set the strength of the DM interactions with the corresponding nucleon.
For a given relativistic theory, reducing the DM--nucleon interaction into its nonrelativistic counterpart generically yields a linear combination of operators.
For example, $\mathcal{O}_1$ through $\mathcal{O}_{11}$ are associated with interactions that occur through the exchange of a heavy spin-0 or spin-1 mediator in a relativistic theory, and the coupling coefficients as written may depend on factors of $|\vec{q}|^2$.
In this study, we avoid choosing specific underlying theories to maintain generality and thus work directly with the operators in Eq.~\eqref{eq:operators}, treating the coupling coefficients as constants.
As detailed in Sec.~\ref{sec:cross-sections}, by considering individual operators, we can study the cosmological effects of DM--nucleon interactions in a systematic manner and place conservative upper limits on each individual coupling coefficient (neglecting operator interference, which we discuss at the end of this section).
We previously constrained interactions via $\mathcal{O}_1$ and $\mathcal{O}_4$ (referred to as the standard ``spin-independent'' and ``spin-dependent'' interactions), corresponding to the simplest case of $n$=0~\cite{Gluscevic:2017ywp}.
We expand upon that work by investigating the remaining 12 operators.
Constraining full relativistic theories is left for future work.

Thus far, we have focused on DM scattering with individual nucleons.
For the purposes of computing signals in direct-detection experiments, these nucleons are embedded in atomic nuclei, and the full response of the nucleus must be treated appropriately.
In a cosmological setting, we are interested in hydrogen and helium nuclei, which dominate the energy density of baryons in the Universe.
Scattering with helium was either neglected or treated too simplistically in previous literature~\cite{Chen:2002yh,Sigurdson:2004zp,Dvorkin:2013cea}; here, we incorporate scattering with helium, accounting for the velocity dependence that arises from its composite nature.
We use the label $B$$\in$$\{p,\textrm{He}\}$ to denote the baryonic species---either the proton or the helium nucleus.
The Hermitian observable $\vec{S}_N$ becomes $\vec{S}_B$ and $\vec{v}^\perp$ becomes $\vec{v}^\perp_B$ (which depends on the reduced mass of the DM--$B$ system, $\mu_{\chi B}$).

\begin{table}[t]
  \begin{minipage}[b]{0.45\linewidth}
    \begin{tabular}{|l|c|c|c|c|c|c|}
      \hline
      Element & $S_B$ & $k$ & $\tau$ & $\tau^\prime$
      & $\mathcal{W}_{B,k}^{\tau\tau^\prime}$ \\
      \hline
      Proton     & $1/2$ & $M$ & 0 & 0 & 1/4 \\
                 &       & $M$ & 0 & 1 & 1/4 \\
                 &       & $M$ & 1 & 0 & 1/4 \\
                 &       & $M$ & 1 & 1 & 1/4 \\
                 &       & $\Sigma^{\prime\prime}$ & 0 & 0 & 1/4 \\
                 &       & $\Sigma^{\prime\prime}$ & 0 & 1 & 1/4 \\
                 &       & $\Sigma^{\prime\prime}$ & 1 & 0 & 1/4 \\
                 &       & $\Sigma^{\prime\prime}$ & 1 & 1 & 1/4 \\
                 &       & $\Sigma^{\prime}$ & 0 & 0 & 1/2 \\
                 &       & $\Sigma^{\prime}$ & 0 & 1 & 1/2 \\
                 &       & $\Sigma^{\prime}$ & 1 & 0 & 1/2 \\
                 &       & $\Sigma^{\prime}$ & 1 & 1 & 1/2 \\
      \hline
      Helium     &   $0$ & $M$ & 0 & 0 & 4 \\
      \hline
    \end{tabular}
    \caption{Terms in the nuclear response functions for each baryonic species $B$.
      The column $S_B$ labels the spin of $B$, $k$ labels the response type, $\tau$ and $\tau^\prime$ label the isospins, and $\mathcal{W}_{B,k}^{\tau\tau^\prime}$ is the numerical coefficient obtained from Ref.~\cite{Catena:2015uha} with an additional factor of $4\pi/(2S_B +1)$.}
    \label{tab:Wresponse}
  \end{minipage}
  \hfill
  \begin{minipage}[b]{0.45\linewidth}
    \begin{tabular}{|c|c|c|c|c|c|}
      \hline
      $k$ & $i$ & $j$ & $\alpha_{ij}$ & $\beta_{ij}$
      & $\mathcal{R}_{k,ij}^{\tau\tau^\prime}$ \\
      \hline
      $M$                    & 1  & 1  & 0 & 0 & $1$ \\
      $M$                    & 5  & 5  & 1 & 1 & $S_\chi(S_\chi + 1)/3$ \\
      $M$                    & 8  & 8  & 1 & 0 & $S_\chi(S_\chi + 1)/3$ \\
      $M$                    & 11 & 11 & 0 & 1 & $S_\chi(S_\chi + 1)/3$ \\
      $\Sigma^{\prime\prime}$  & 10 & 10 & 0 & 1 & $1/4$ \\
      $\Sigma^{\prime\prime}$  & 4  & 4  & 0 & 0 & $S_\chi(S_\chi + 1)/12$ \\
      $\Sigma^{\prime\prime}$  & 4  & 6  & 0 & 1 & $S_\chi(S_\chi + 1)/12$ \\
      $\Sigma^{\prime\prime}$  & 6  & 4  & 0 & 1 & $S_\chi(S_\chi + 1)/12$ \\
      $\Sigma^{\prime\prime}$  & 6  & 6  & 0 & 2 & $S_\chi(S_\chi + 1)/12$ \\
      $\Sigma^{\prime\prime}$  & 12 & 12 & 1 & 0 & $S_\chi(S_\chi + 1)/12$ \\
      $\Sigma^{\prime\prime}$  & 13 & 13 & 1 & 1 & $S_\chi(S_\chi + 1)/12$ \\
      $\Sigma^{\prime}$       & 3  & 3  & 1 & 1 & $1/8$ \\
      $\Sigma^{\prime}$       & 7  & 7  & 1 & 0 & $1/8$ \\
      $\Sigma^{\prime}$       & 4  & 4  & 0 & 0 & $S_\chi(S_\chi + 1)/12$ \\
      $\Sigma^{\prime}$       & 9  & 9  & 0 & 1 & $S_\chi(S_\chi + 1)/12$ \\
      $\Sigma^{\prime}$       & 12 & 12 & 1 & 0 & $S_\chi(S_\chi + 1)/24$ \\
      $\Sigma^{\prime}$       & 15 & 15 & 1 & 2 & $S_\chi(S_\chi + 1)/24$ \\
      $\Sigma^{\prime}$       & 12 & 15 & 1 & 1 & $-S_\chi(S_\chi + 1)/24$ \\
      $\Sigma^{\prime}$       & 15 & 12 & 1 & 1 & $-S_\chi(S_\chi + 1)/24$ \\
      $\Sigma^{\prime}$       & 14 & 14 & 1 & 1 & $S_\chi(S_\chi + 1)/24$ \\
      \hline
    \end{tabular}
    \caption{Terms in the DM response functions.
      The column $k$ labels the response type, $i$ and $j$ label the operators associated with the coupling coefficients, $\alpha_{ij}$ labels powers of $|\vec{v}|_T^{\perp 2}$, $\beta_{ij}$ labels powers of $|\vec{q}|^2/m_N^2$, and $\mathcal{R}_{k,ij}^{\tau\tau^\prime}$ is the numerical coefficient obtained from Ref.~\cite{Anand:2013yka}.}
    \label{tab:Rresponse}
  \end{minipage}
\end{table}

In Refs.~\cite{Fitzpatrick:2012ix,Anand:2013yka}, DM interactions with composite nuclei are approximated as the sum of interactions with the individual nucleons.
Additionally, the nuclear wave functions are taken to have the standard shell-model form, and the underlying single-particle basis is the harmonic oscillator with a parametric size
\begin{equation}
  a_B = \sqrt{41.467/(45 A^{-1/3} - 25 A^{-2/3})}~\fm \ ,
  \label{eq:aB}
\end{equation}
where $A$$>$1 is the atomic number.
The proton is pointlike and thus $a_p$=0.
The effect of compositeness when scattering with a nonpointlike nucleus is encoded in the nuclear response function $W_{B,k}^{\tau\tau^\prime}(y)$, where $y$$\equiv$$(|\vec{q}|a_B/2)^2$.
The index $k$ labels the type of response (not to be confused with the wave number defined in Sec.~\ref{sec:boltzmann}), as in the standard treatment of semileptonic weak interactions~\cite{Donnelly:1979ezn,Serot:1979yk}.
There is also a DM response function $R_k^{\tau\tau^\prime} (\vec{v}_B^{\perp 2},|\vec{q}|^2/m_N^2)$, which incorporates the coupling coefficients of the various operators.
The resulting spin-averaged amplitude squared for the scattering between DM and the baryon $B$ is~\cite{Anand:2013yka}
\begin{equation}
  \Msq_B = \frac{1}{m_v^4} \frac{4\pi}{2S_B + 1} \sum_{\tau, \tau^\prime} \sum_k
  \left(\frac{|\vec{q}|^2}{m_N^2}\right)^{\xi_k}
  R_k^{\tau\tau^\prime}\left(\vec{v}_B^{\perp 2},\frac{|\vec{q}|^2}{m_N^2}\right)
  W_{B,k}^{\tau\tau^\prime}(y) \ ,
  \label{eq:Msq-general}
\end{equation}
where $\xi_k$=0 for $k$$\in$$\{M, \Sigma^{\prime\prime}, \Sigma^{\prime}\}$ and $\xi_k$=1 for $k$$\in$$\{\Phi^{\prime\prime},\Phi^{\prime\prime}M,\widetilde{\Phi}^{\prime},\Delta,\Delta\Sigma^{\prime}\}$.
We have inserted the mass of the weak scale $m_v$$\equiv$$(\sqrt{2}G_F)^{-1/2}$$\approx$$246~\GeV$ to render the coupling coefficients dimensionless.\footnote{This mass scale is an arbitrary normalization, and it does not impact the numerical value for the constraints on the cross sections reported in Sec.~\ref{sec:planck}.}
We refer the reader to Ref.~\cite{Anand:2013yka} for the full expression for $R_k^{\tau\tau^\prime}$ and to Refs.~\cite{Fitzpatrick:2012ix,Catena:2015uha} for the form of $W_{B,k}^{\tau\tau^\prime}$ for various elements.

In the context of cosmology, we need only consider scattering on protons (with response types $M$, $\Sigma^{\prime\prime}$, and $\Sigma^{\prime}$) and helium nuclei (with response type $M$).
It will also be useful to isolate the velocity and angular dependencies in Eq.~\eqref{eq:Msq-general}.
For this purpose, we first define the dimensionless quantity $x$$\equiv$$|\vec{q}|^2/|\vec{q}|^2_\textrm{max}$ and make the following substitutions:
\begin{equation}
  \frac{|\vec{q}|^2}{m_N^2} = x v^2 \left(\frac{2\mu_{\chi B}}{m_N}\right)^2 \ , \qquad
  |\vec{v}_B^\perp|^2 = v^2(1-x) \ , \qquad
  y = x v^2(\mu_{\chi B} a_B)^2 \ .
\end{equation}
Next, we express the DM response function as a product of a numerical coefficient $\mathcal{R}_{k,ij}$, the coupling coefficients of $\mathcal{O}_i$ and $\mathcal{O}_j$, and powers of $|\vec{v}|_B^{\perp 2}$ and $(|\vec{q}|/m_N)^2$, such that
\begin{equation}
  R_k^{\tau\tau^\prime}
  = \sum_{i,j}c_i^\tau c_j^{\tau^\prime} \mathcal{R}_{k,ij}
  \left(|\vec{v}|_B^{\perp 2}\right)^{\alpha_{ij}}
  \left(\frac{|\vec{q}|^2}{m_N^2}\right)^{\beta_{ij}}
  = \sum_{i,j} c_i^\tau c_j^{\tau^\prime} \mathcal{R}_{k,ij}
  \left(\frac{2\mu_{\chi B}}{m_N}\right)^{2\beta_{ij}}
  v^{2(\alpha_{ij}+\beta_{ij})} (1-x)^{\alpha_{ij}} x^{\beta_{ij}} \ .
\end{equation}
The values of $\alpha_{ij}$, $\beta_{ij}$, and $\mathcal{R}_{k,ij}$ are listed in Table~\ref{tab:Rresponse}.\footnote{For the most generic form of $R_k^{\tau\tau^\prime}$, the powers of $|\vec{v}|_B^{\perp 2}$ and $(|\vec{q}|/m_N)^2$ are determined not only by the operators, but also by the response type $k$ involved; thus, they should be written as $\alpha_{k,ij}$ and $\beta_{k,ij}$. As evident in Table~\ref{tab:Rresponse}, however, the values of $\alpha_{ij}$ and $\beta_{ij}$ are the same for fixed $i$ and $j$ across all $k$ (with nonzero $\mathcal{R}_{k,ij}$) under consideration. Hence, we may drop the $k$ index for the subset of response types that are relevant for cosmology.}
We factorize the nuclear response function in a similar manner to obtain
\begin{equation}
  W_{B,k}^{\tau\tau^\prime} = \frac{2S_B+1}{4\pi} \mathcal{W}_{B,k}^{\tau\tau'} e^{-2y}
  = \frac{2S_B+1}{4\pi} \mathcal{W}_{B,k}^{\tau\tau'} e^{-2x v^2 (\mu_{\chi B}a_B)^2} \ ,
\end{equation}
where the values of the numerical factors $\mathcal{W}_{B,k}^{\tau\tau'}$ are listed in Table~\ref{tab:Wresponse}.
This expression is valid for hydrogen and helium,\footnote{For arbitrary nuclei, the expression for the nuclear response function is a polynomial in $y$ multiplied by $e^{-2y}$. However, the nuclear response is constant for the proton, and the polynomial in $y$ is simply a constant for helium.} noting that the velocity dependence in the exponential is removed for hydrogen when setting $a_p$=0.
We may now recast Eq.~\eqref{eq:Msq-general} as
\begin{align}
  \Msq_B = \frac{1}{m_v^4} \sum_{i,j}\sum_{\tau, \tau^\prime} \sum_k
  c_i^\tau c_j^{\tau^\prime} \mathcal{R}_{k,ij} \mathcal{W}_{B,k}^{\tau\tau'}
  \left(\frac{2\mu_{\chi B}}{m_N}\right)^{2\beta_{ij}} v^{2(\alpha_{ij}+\beta_{ij})}
  (1-x)^{\alpha_{ij}} x^{\beta_{ij}} e^{-2x v^2 (\mu_{\chi B}a_B)^2} \ .
  \label{eq:Mhhe}
\end{align}
It is possible to have nonzero terms for which $i$$\neq$$j$, indicating interference between operators; specifically, as seen in Table~\ref{tab:Rresponse}, there is interference between $\mathcal{O}_4$ and $\mathcal{O}_6$ and between $\mathcal{O}_{12}$ and $\mathcal{O}_{15}$.
However, when handling a single operator at a time, the interference between operators plays no role.
There is constructive interference between $\mathcal{O}_4$ and $\mathcal{O}_6$; hence, we expect the upper limits on the coupling coefficients in Sec.~\ref{sec:planck} to be conservative: if another operator contributes to a signal, then the upper limit can, in principle, be made stronger.
For $\mathcal{O}_{12}$ and $\mathcal{O}_{15}$, there is destructive interference, and we do not expect to obtain the most conservative upper limits on the coupling coefficients for the individual analyses of these two operators.

\subsection{Cross Sections}
\label{sec:cross-sections}

The differential cross section, written as a function of $x$, for DM scattering with a baryon $B$ is
\begin{equation}
  \frac{d\sigma_B}{dx} = \frac{\mu_{\chi B}^2}{\pi} \Msq_B \ .
  \label{eq:diffcs}
\end{equation}
In Sec.~\ref{sec:planck}, we express our cosmological constraints in terms of the total cross section $\sigma_B(v)$ for scattering with protons, found by integrating Eq.~\eqref{eq:diffcs} over $x$:
\begin{align}
  \sigma_B(v)
  = \int_0^1 \frac{d\sigma_B}{dx}\; dx
  &= \sum_{i,j} \widetilde{\sigma}_{B}^{(ij)} v^{2(\alpha_{ij}+\beta_{ij})}
  {}_1F_1 (1+\beta_{ij}, 2+\alpha_{ij}+\beta_{ij};-2\mu_{\chi B}^2 a_B^2 v^2),
  \label{eq:cs1}
\end{align}
where
\begin{equation}
  \widetilde{\sigma}_{B}^{(ij)} \equiv \frac{\mu_{\chi B}^2}{\pi m_v^4}
  \left(\frac{2\mu_{\chi B}}{m_N}\right)^{2\beta_{ij}}
  \sum_{\tau\tau^\prime} \sum_k c_i^\tau c_j^{\tau^\prime}
  \mathcal{R}_{k,ij} \mathcal{W}_{B,k}^{\tau\tau^\prime} \ ,
  \label{eq:cs-prefactor}
\end{equation}
and ${}_1F_1$ is the confluent hypergeometric function of the first kind.
However, as mentioned in Sec.~\ref{sec:cosmo}, the key quantity that affects cosmological observables is the momentum-transfer cross section $\sigma_{\textrm{MT},B}(v)$, which weights the differential cross section by a factor of $(1-\cos\theta)$=$2x$ to preferentially pick out scattering processes with large momentum transfer:
\begin{align}
  \sigma_{\textrm{MT},B}(v)
  = 2 \int_0^1 x \frac{d\sigma_B}{dx}\; dx
  &= \sum_{i,j} \widetilde{\sigma}_{B}^{(ij)} v^{2(\alpha_{ij}+\beta_{ij})}
  \frac{2(1+\beta_{ij})}{2+\alpha_{ij}+\beta_{ij}}
  {}_1F_1 (2+\beta_{ij}, 3+\alpha_{ij}+\beta_{ij};-2\mu_{\chi B}^2 a_B^2 v^2) \ .
  \label{eq:cs2}
\end{align}
The velocity dependence of $\sigma_B(v)$ and $\sigma_{\textrm{MT},B}(v)$ may arise either from the structure of the interaction itself [\textit{i.e.}, from the DM response function, captured in the power-law index $2(\alpha_{ij}$$+$$\beta_{ij})$] or from the composite nature of the nucleus (\textit{i.e.}, from the nuclear response function, captured in the ${}_1F_1$ function).
The latter is nontrivial only for helium in this study, while for hydrogen it evaluates to 1, leaving the cross section and momentum-transfer cross section with a simple power-law velocity dependence.

It is important to note the various velocity dependencies for discerning observational signatures of different interactions; as described in Sec.~\ref{sec:boltzmann}, the velocity dependence influences the time evolution of the momentum transfer between DM and baryon fluids, as well as the time of their thermal decoupling.
The rate of momentum transfer in turn controls the relative size of the effect of scattering on different density modes in the early Universe, producing different signatures in the CMB and the matter power spectrum. We discuss the effect of scattering on cosmological observables in detail in Sec.~\ref{sec:observables}.

In this study, we focus on DM interactions with protons and neglect interactions with neutrons embedded in helium nuclei; we can thus replace all coupling coefficients labeled with isospin by the coupling coefficients with protons, which we simply denote as $c_i$, dropping the superscript ``$(p)$'' henceforth.
Since we also focus on a single operator at a time, Eqs.~\eqref{eq:cs1}, \eqref{eq:cs-prefactor}, and \eqref{eq:cs2} simplify to
\begin{align}
  \sigma_B^{(i)}(v)
  &= \widetilde{\sigma}_{B}^{(i)} v^n
  {}_1F_1 (1+\beta, 2+\alpha+\beta;-2\mu_{\chi B}^2 a_B^2 v^2) \nonumber\\
  \sigma_{\textrm{MT},B}^{(i)}(v)
  &= \widetilde{\sigma}_{B}^{(i)} v^n
  \frac{2(1+\beta)}{2 + \alpha + \beta}
  {}_1F_1 (2+\beta, 3 + \alpha + \beta;-2\mu_{\chi B}^2 a_B^2 v^2) \nonumber\\
  \widetilde{\sigma}_{B}^{(i)} &= \frac{\mu_{\chi B}^2 c_i^2}{\pi m_v^4}
  \left(\frac{2\mu_{\chi B}}{m_N}\right)^{2\beta}
  \sum_{\tau\tau^\prime} \sum_k
  \mathcal{R}_{k,ij} \mathcal{W}_{B,k}^{\tau\tau^\prime} \ ,
  \label{eq:cs-simple}
\end{align}
where we eliminated the indices on $\alpha$ and $\beta$, with the understanding that they correspond to $\alpha_{ii}$ and $\beta_{ii}$, respectively.
Additionally, we define the quantity $n$$\equiv$$2(\alpha + \beta)$ to emphasize that the cross section has a single power-law index of the relative velocity for a given operator.
As shown in Table~\ref{tab:Rresponse}, the operators cover $n$$\in$$\{0,2,4,6\}$.
However, the entire phenomenology of these different interactions does not reduce to the choice of $n$; namely, we find in Sec.~\ref{sec:planck} that different operators with the same $n$ but a different set of $(\alpha,\beta)$ yield different mass dependencies for the cosmological constraints on the coupling coefficients.

\section{Cosmological setting}
\label{sec:cosmo}

In this section, we embed the EFT formalism into a cosmological setting.
We start in Sec.~\ref{sec:boltzmann} by reviewing the modified system of Boltzmann equations that describe the evolution of cosmological perturbations in the presence of DM--proton interactions.
We make the necessary modifications to all relevant background quantities, such as the DM temperature and the heat and momentum transfer rates between the DM and baryon fluids.
In Sec.~\ref{sec:observables}, using our implementation of these equations in the code \texttt{CLASS}~\cite{Blas:2011rf}, we illustrate the effect of interactions on cosmological observables.

\subsection{Boltzmann equations}
\label{sec:boltzmann}

We work in Fourier space to express the evolution of the DM and baryon density fluctuations, $\delta_\chi$ and $\delta_b$, and their velocity divergences, $\theta_\chi$ and $\theta_b$, respectively, as
\begin{align}
  \dot{\delta}_\chi &= -\theta_\chi - \frac{\dot{h}}{2}
  &
  \dot{\theta}_\chi &= -\frac{\dot{a}}{a}\theta_\chi + c_\chi^2 k^2 \delta_\chi
  + R_\chi (\theta_b - \theta_\chi)
  \nonumber\\
  \dot{\delta}_b    &= -\theta_b    - \frac{\dot{h}}{2}
  &
  \dot{\theta}_b    &= -\frac{\dot{a}}{a}\theta_b   + c_b^2 k^2 \delta_b
  + R_\gamma (\theta_\gamma - \theta_b)
  + \frac{\rho_\chi}{\rho_b} R_\chi (\theta_\chi - \theta_b)
  \label{eq:fluct}
\end{align}
in the synchronous gauge, where $k$ is the wave number of a given Fourier mode (not to be confused with the index denoting DM response function types of Sec.~\ref{sec:EFT}); $a$ is the scale factor; $h$ is the trace of the scalar metric perturbation~\cite{Ma:1995ey}; ${c}_b$ and $c_\chi$ are the speeds of sound in the two fluids~\cite{Ma:1995ey}; and $\rho_b$ and $\rho_\chi$ are their respective energy densities.
The overdot represents a derivative with respect to conformal time, and the subscript $\gamma$ pertains to photons.
The rate coefficients $R_\gamma$ and $R_\chi$ arise from scattering processes that change particle momenta, resulting in a drag force that affects the evolution of $\theta_b$ and $\theta_\chi$.
The standard term $R_\gamma$ is associated with Compton scattering~\cite{Ma:1995ey}; the term $R_\chi$ arises from DM--baryon scattering~\cite{Chen:2002yh,Sigurdson:2004zp}.

In a single collision, the momentum of the DM particle changes by~\cite{Dvorkin:2013cea}
\begin{equation}
  |\Delta\vec{p}_\chi| = \frac{m_\chi m_B}{m_\chi + m_B} |\vec{v}_\chi - \vec{v}_B|
  \left(\hat{n} - \frac{\vec{v}_\chi - \vec{v}_B}{|\vec{v}_\chi - \vec{v}_B|}\right) \ ,
\end{equation}
where $m_B$ is the mass of baryon $B$, $\hat{n}$ is the direction of the scattered DM particle in the center-of-mass frame, $\vec{v}_\chi$ is the DM velocity, and $\vec{v}_B$ is the velocity of baryon $B$.
The relative velocity is $\vec{v}$=$\vec{v}_\chi - \vec{v}_B$.
In line with Sec.~\ref{sec:EFT}, quantities labeled with ``$B$'' indicate a particular baryonic species and those labeled with ``$b$'' refer to the baryon fluid as a whole.
We assume phase space distribution functions of the form
\begin{align}
  f_\chi(\vec{v}_\chi) &= \frac{n_\chi}{(2\pi)^{3/2} \bar{v}_\chi^{3/2}}
  \exp\left[-(\vec{v}_\chi-\vec{V}_\chi)^2/(2\bar{v}_\chi^2)\right] \\
  f_B(\vec{v}_B) &= \frac{n_B}{(2\pi)^{3/2} \bar{v}_B^{3/2}}
  \exp\left[-(\vec{v}_B-\vec{V}_b)^2/(2\bar{v}_B^2)\right] \ ,
\end{align}
where $\vec{V}_\chi$ and $\vec{V}_b$ are the peculiar velocities of the DM and baryon fluids, and $\bar{v}_\chi^2$=$T_\chi / m_\chi$ and $\bar{v}_B^2$=$T_b / m_B$ are the DM and baryon velocity dispersions, respectively.
The resulting drag force per unit mass, or drag acceleration, on the DM fluid is given by
\begin{align}
  \frac{d\vec{V}_\chi}{dt}
  &= -\frac{1}{m_\chi} \sum_B \int d^3v_\chi d^3v_B f_\chi(\vec{v}_\chi) f_B(\vec{v}_B)
  \int \frac{d\sigma}{d\Omega} |\vec{v}_\chi - \vec{v}_B|
  |\Delta\vec{p}_\chi| \nonumber\\
  &= -\sum_B \frac{\rho_B}{m_\chi+m_B}
  \frac{1}{(2\pi)^{3/2}} \frac{1}{(\bar{v}_B^2 + \bar{v}_\chi^2)^{3/2}}
  \int d^3v\, \vec{v} \left[\sigma_{\textrm{MT},B}^{(i)}(v) v\right]
  \exp\left\{-\frac{[\vec{v}-(\vec{V}_\chi-\vec{V}_b)]^2}
                  {2(\bar{v}_B^2 + \bar{v}_\chi^2)}\right\} \ .
\end{align}
In the limit where the DM--baryon relative bulk velocity is small compared to thermal velocity,\footnote{The condition $(\vec{V}_\chi - \vec{V}_b)^2 \ll (\bar{v}_\chi^2 + \bar{v}_B^2)$ is satisfied for interactions that couple DM to protons at early times only; in all the cases we consider, DM--proton decoupling occurs at $z$$>$$10^4$, as shown in Figure~\ref{fig:RT}.} the final expression for the rate coefficient at a given redshift is
\begin{align}
  R_\chi = a\rho_b \sum_B
  &\frac{Y_B}{m_\chi+m_B}
  \widetilde{\sigma}_B^{(i)} \frac{2(1+\beta)}{2+\alpha+\beta} \mathcal{N}_n
  \left(\frac{T_b}{m_B}+\frac{T_\chi}{m_\chi}\right)^{1/2+\alpha + \beta} \nonumber \\
  & \times
  \left[1+(2\mu_{\chi B} a_B)^2 \left(\frac{T_b}{m_B}+\frac{T_\chi}{m_\chi}\right)\right]^{-(2+\beta)} \ ,
  \label{eq:rate}
\end{align}
where $Y_B$ is the mass fraction of the baryon $B$; $\mathcal{N}_n$$\equiv$$2^{(5+n)/2} \Gamma(3+n/2) / (3\sqrt{\pi})$ is a numerical factor; and $T_b$ and $T_\chi$ are the baryon and DM temperatures, respectively.

Finally, since we are interested in sub-GeV DM, we cannot neglect terms with $T_\chi$ in the above equations, as was the approach in Ref.~\cite{Dvorkin:2013cea}.
We thus track the coupled evolution of the DM and baryon temperatures, given by
\begin{align}
  \dot{T}_\chi &= -2\frac{\dot{a}}{a}T_\chi
  + 2 R^\prime_\chi (T_b - T_\chi) \nonumber\\
  \dot{T}_b &= -2\frac{\dot{a}}{a}T_b
  + \frac{2\mu_b}{m_\chi} \frac{\rho_\chi}{\rho_b}R^\prime_\chi (T_\chi - T_b)
  + \frac{2\mu_b}{m_e} R_\gamma (T_\gamma - T_b) \ ,
  \label{eq:temp}
\end{align}
where $m_e$ is the electron mass; $\mu_b$$\approx$$m_\textrm{H} (n_\textrm{H}+4n_\textrm{He})/(n_\textrm{H}+n_\textrm{He}+n_e)$ is the mean molecular weight of the baryons; $n_\textrm{H}$ and $n_\textrm{He}$ are the number densities of protons and helium nuclei, respectively; and the heat-exchange rate coefficient $R_\chi^\prime$ is given by Eq.~\eqref{eq:rate}, but with an additional factor of $m_\chi/(m_\chi + m_B)$ multiplying each summand.

\subsection{Effect of Scattering on Cosmological Observables}
\label{sec:observables}

\begin{figure*}[t]
\includegraphics[height=0.33\textwidth]{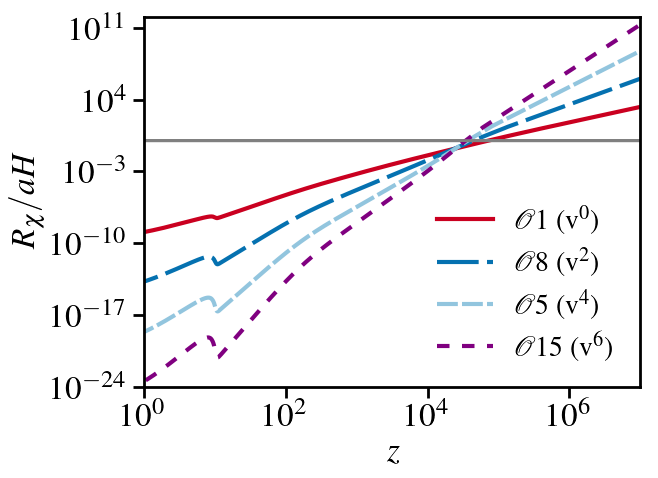}
\hspace{0.8cm}
\includegraphics[height=0.33\textwidth]{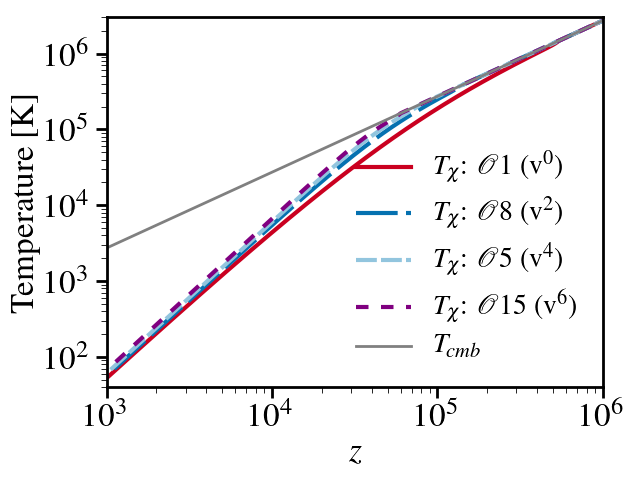}
\caption{\textbf{[Left]} Redshift evolution of the coefficient $R_\chi$, defined in Eq.~\eqref{eq:rate}, which quantifies the rate of momentum exchange between the DM and baryon fluids. It is normalized to the Hubble expansion rate. \textbf{[Right]} Redshift evolution of the DM temperature $T_\chi$. The CMB temperature (thin gray line) is also plotted for reference. \textbf{[Both]} These background quantities are shown for a subset of DM--proton interactions with various relative velocity scalings of the cross section (indicated in the legend). We fix the DM spin to $S_\chi$=$1/2$ and the DM particle mass to $m_\chi$=$1~\GeV$, and set the coupling coefficients to their respective 2$\sigma$ upper limits reported in Table~\ref{tab:alllims}, keeping other cosmological parameters at their best-fit \textit{Planck} 2015 values.}
\label{fig:RT}
\end{figure*}

We have modified the linear Boltzmann solver \texttt{CLASS} to implement the Boltzmann equations given in Sec.~\ref{sec:boltzmann}, incorporating the effect of DM--proton scattering on the evolution of cosmological perturbations, for all operators in Eq.~\eqref{eq:operators}.
Figure~\ref{fig:RT} shows the relevant background quantities, $T_\chi$ and $R_\chi$, as functions of redshift for a subset of operators that correspond to interactions whose cross sections scale with different powers of velocity (labeled as $v^n$ in the legends of the figure).
To illustrate the evolution of these quantities, we fix $S_\chi$=$1/2$ and $m_\chi$=$1~\GeV$, set the coupling coefficients to their respective 2$\sigma$ upper limits reported in Table~\ref{tab:alllims}, and keep other cosmological parameters at their best-fit \textit{Planck} 2015 values~\cite{Ade:2015xua}.
A stronger velocity dependence (larger $n$) leads to more momentum transfer at early times and to a later thermal decoupling time of the DM and baryon fluids.
This difference in the evolution of $R_\chi$ determines the relative size of the effect of DM--baryon scattering on different density perturbation modes, traceable through cosmological observables such as the CMB power spectra $C_\ell$ and the three-dimensional matter power spectrum $P(k)$.

\begin{figure*}[t]
\includegraphics[height=0.32\textwidth]{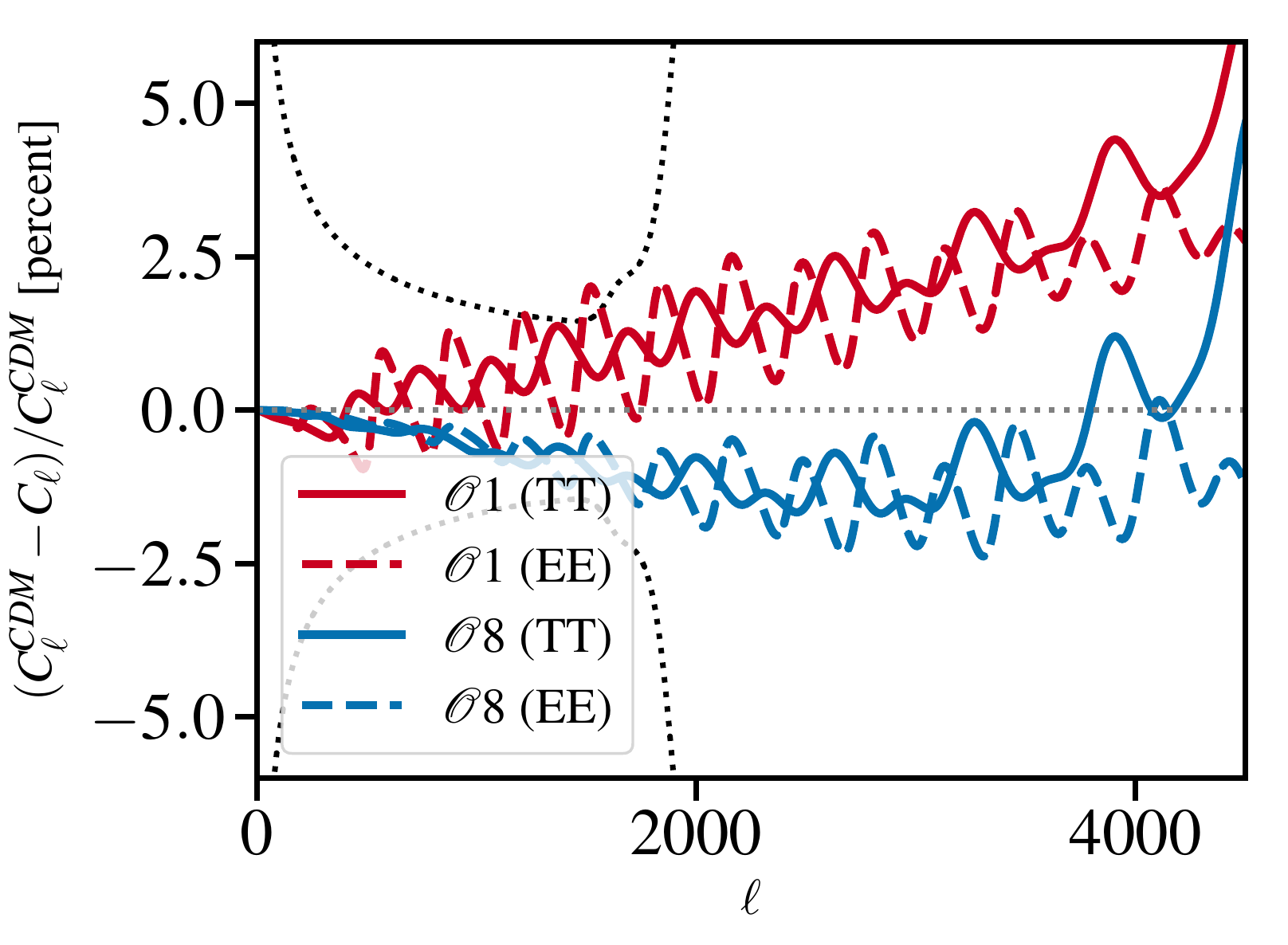}
\hspace{0.8cm}
\includegraphics[height=0.32\textwidth]{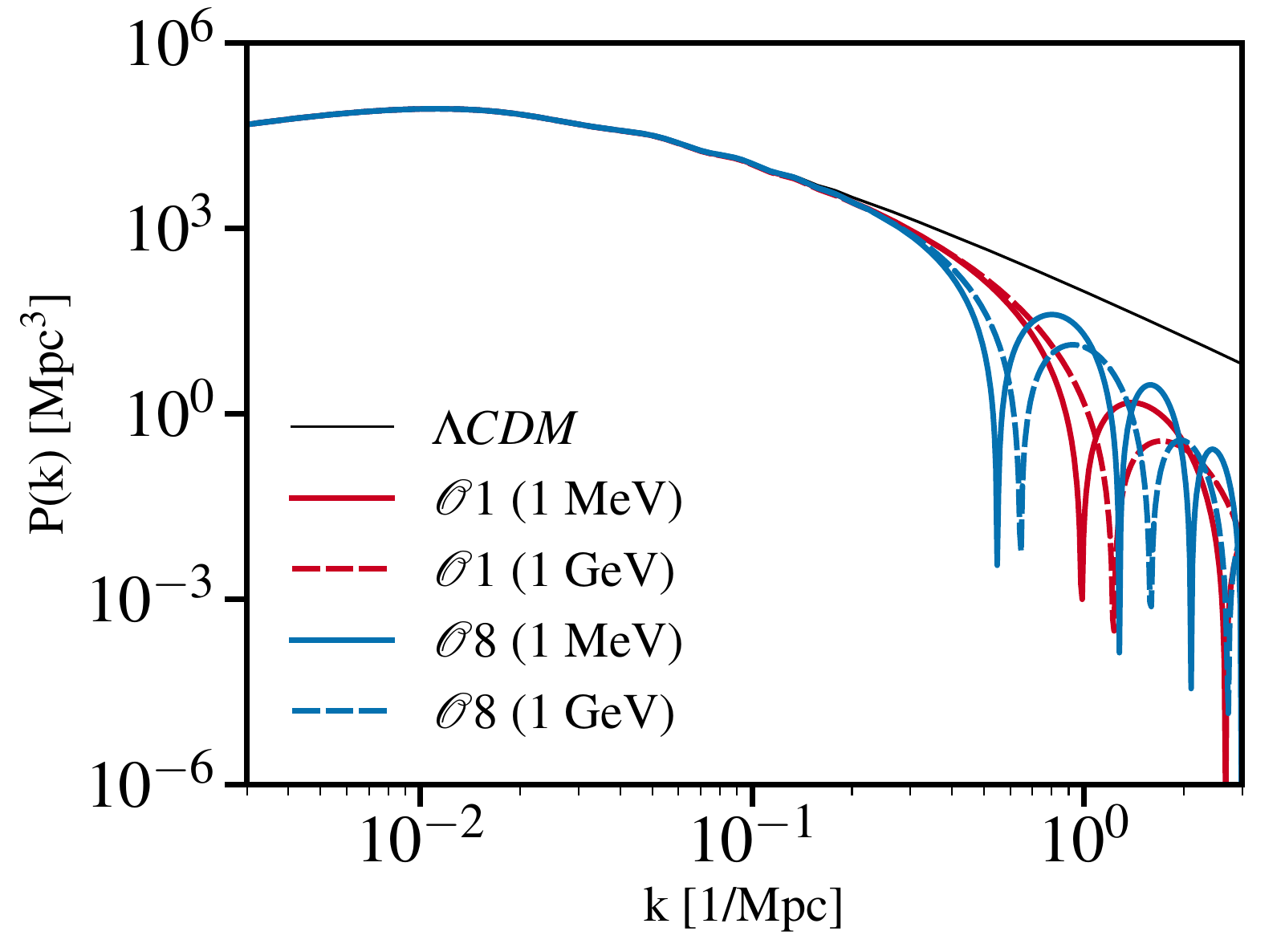}
\caption{\textbf{[Left]} Percent difference between the CMB temperature (solid) and $EE$ polarization (dashed) power spectra, computed for the standard ``CDM'' scenario with no DM interactions and for the scenario with DM--proton coupling, for DM mass $m_\chi$=$1~\GeV$. Power spectra are shown for scattering through $\mathcal{O}_1$ (top set of red lines) and $\mathcal{O}_8$ (bottom set of blue lines), as examples of velocity-independent and velocity-dependent scattering, respectively. The dotted black lines roughly represent the \textit{Planck} 2$\sigma$ error bars, binned in multipoles with $\Delta\ell$=50. Positive differences indicate a suppression of power (with respect to ``CDM''), while negative differences indicate an enhancement of power. \textbf{[Right]} Matter power spectra for the scenario of no DM scattering (black) compared to the scenarios where there is scattering through $\mathcal{O}_1$ (red) and $\mathcal{O}_8$ (blue), for two different DM particle masses: $1~\GeV$ (dashed) and $1~\MeV$ (solid). \textbf{[Both]} For the cases with interactions, we fix the DM spin to $S_\chi$=$1/2$ and set the coupling coefficients to their respective 2$\sigma$ upper limits reported in Table~\ref{tab:alllims}, keeping other cosmological parameters at their best-fit \textit{Planck} 2015 values. \label{fig:res}}
\end{figure*}

In Figure~\ref{fig:res}, we illustrate the effect of scattering on these observables for $\mathcal O_{1}$ (with $n$=0) and $\mathcal O_{8}$ (with $n$=2) by comparing to the standard ``CDM'' case with no DM--proton interactions.
For DM--proton scattering, we fix $S_\chi$=$1/2$ and set the coupling coefficients to their 2$\sigma$ upper limits reported in Table~\ref{tab:alllims}, keeping all other cosmological parameters at their best-fit \textit{Planck} 2015 values, as in Figure~\ref{fig:RT}.

For the left panel of Figure~\ref{fig:res}, we set $m_\chi$=$1~\GeV$.
To get a sense for the range of $\ell$ multipoles in which the scattering signal is most prominent, we show the percent difference between the ``CDM'' case and the cases with DM--proton scattering.
As was previously noted in Refs.~\cite{Chen:2002yh,Sigurdson:2004zp}, on large angular scales in the CMB, the coupling of DM to baryons presents similarly to baryon loading: the tight coupling with DM particles at early times effectively increases the total mass of the baryons, enhancing power at low multipoles (\textit{c.f.} negative percent differences).
On small scales, the drag force between baryons and DM dissipates the momentum of the baryon--photon fluid, damping baryon acoustic oscillations and suppressing power more for modes that enter the cosmological horizon earlier (roughly corresponding to larger multipoles).
Furthermore, acoustic peaks shift to smaller angular scales as a result of the decrease in the speed of sound in primordial plasma.
Since the $EE$ polarization spectra display sharper oscillatory features in multipole space compared to $TT$, the shift produces higher-amplitude oscillations seen in this difference plot.

In the right panel of Figure~\ref{fig:res}, we show the matter power spectrum today, comparing the ``CDM'' case to the cases with DM--proton scattering for $m_\chi$=$1~\MeV$ and $m_\chi$=$1~\GeV$.
When DM couples to protons, $P(k)$ exhibits oscillatory suppressions at large $k$ due to DM tightly following the behavior of the baryon--photon fluid and undergoing ``dark oscillations'' at early times, when the corresponding modes enter the horizon.
For heavier DM particles, the oscillations are shifted toward larger $k$, indicating earlier thermal decoupling from baryons; at a fixed DM energy density, increasing the DM mass results in a lower particle number density and thus a reduced interaction rate to maintain thermal equilibrium.

With the ability to compute the power spectra for a given cosmology, we proceed to search for signals consistent with DM--baryon interactions within the data from the \textit{Planck} satellite.

\section{Numerical results}
\label{sec:planck}

We perform a joint likelihood analysis of \textit{Planck} 2015 temperature, polarization, and lensing anisotropy measurements, using the Planck Likelihood Code v2.0 (\texttt{Clik/Plik}) \cite{2016A&A...594A..11P,2016A&A...594A...1P} to search for evidence of DM--proton scattering in the early Universe.
For high multipoles,  we use the nuisance-marginalized joint $TT$, $TE$, $EE$ likelihood (\texttt{Clik/Plik} {lite}), with $\ell$=30--2508 for $TT$ and $\ell$=30--1996 for $TE$ and $EE$.
Using \texttt{Clik/Plik} {lite} over the full likelihood requires substantially fewer computational resources, and we find no significant effect on the derived values of cosmological parameters and constraints.\footnote{\textit{Planck} high-multipole polarization could have systematic issues~\cite{2016A&A...594A..11P,2016A&A...594A...1P} that may affect cosmological parameter estimation; however, excluding $\ell$$\geq$30 polarization would degrade our reported constraints on DM interactions by less than $8\%$.}
The lensing likelihood we use contains SMICA map-based lensing reconstruction for multipoles in the range $\ell$=40--400.

To sample the cosmological parameter space, we employ \texttt{MontePython}~\cite{Audren:2012wb} with the \texttt{PyMultinest}~\cite{2014A&A...564A.125B} implementation of the nested sampling algorithm~\cite{Feroz:2007kg,Feroz:2008xx,Feroz:2013hea}.
We have verified that our pipeline recovers the published $\Lambda$CDM parameter values and constraints~\cite{Ade:2015xua} to within $0.14\sigma$, in the limit of vanishing DM coupling coefficients.
Since the convergence criteria for nested likelihood sampling are not clearly defined (in contrast to the case of MCMC methods), we check for convergence in our samples by varying the number of live points passed to \texttt{PyMultinest} and ensuring that the inferred constraints do not change by more than an order of percent in numerical value between different sampling runs.
We assume that baryons, photons, and DM are tightly coupled and in thermal equilibrium at the start of the integration of the Boltzmann equations; this condition is satisfied for all interactions at their $1\sigma$ and $2\sigma$ exclusion limits in Table~\ref{tab:alllims}.
Additionally, we consider only a flat geometry.

We perform a likelihood analysis for each individual interaction operator, repeating the fitting procedure for the following fixed values of DM mass: $m_\chi$$\in$$\{15~\keV$, $1~\MeV$, $1~\GeV$, $1~\TeV\}$.
Thus, in addition to the six standard $\Lambda$CDM parameters (baryon density $\Omega_bh^2$, DM density $\Omega_\chi h^2$, Hubble parameter $h$, reionization optical depth $\tau$, amplitude of the scalar perturbations $A_s$, and scalar spectral index $n_s$), each sampling run also includes the coupling coefficient $c_i$ for $\mathcal{O}_i$ as a free fitting parameter; we assume wide flat priors for all parameters.
We obtain the 68\% and 95\% confidence-level exclusion limits for the coupling coefficients $c_i$ from their posterior probability distributions, and we convert these limits to ones for the corresponding interaction cross section using Eq.~\eqref{eq:cs-simple}.

\begin{table}[t]
\begin{tabular}{ |c|c|c|c|c| }
\hline
\textbf{Operator} & \multicolumn{4}{|c|}{\textbf{DM Mass}} \\
\cline{2-5}
\textbf{[$n$ ($\alpha$+$\beta$)]} & \textbf{15 keV} & \textbf{1 MeV} & \textbf{1 GeV} & \textbf{1 TeV}\\
\hline
$\mathcal{O}_{1}$ [0 (0+0)] & 2.9e-27 (8.8e-27) & 9.1e-27 (2.6e-26) & 4.9e-26 (1.5e-25) & 4.7e-24 (1.4e-23)\\
\hline
$\mathcal{O}_{3}$ [4 (1+1)] & 2.3e-33 (5.7e-33) & 1.4e-29 (3.8e-29) & 6.5e-24 (1.9e-23) & 9.6e-21 (3.4e-20)\\
\hline
$\mathcal{O}_{4}$ [0 (0+0)] & 3.7e-27 (1.2e-26) & 1.1e-26 (3.3e-26) & 9.3e-26 (2.9e-25) & 5.6e-23 (1.7e-22)\\
\hline
$\mathcal{O}_{5}$ [4 (1+1)] & 1.9e-33 (4.6e-33) & 1.1e-29 (3.0e-29) & 4.0e-24 (1.2e-23) & 8.6e-22 (2.7e-21)\\
\hline
$\mathcal{O}_{6}$ [4 (0+2)] & 1.5e-33 (3.8e-33) & 9.6e-30 (2.4e-29) & 4.5e-24 (1.3e-23) & 6.4e-21 (2.1e-20)\\
\hline
$\mathcal{O}_{7}$ [2 (1+0)] & 1.0e-29 (2.8e-29) & 1.1e-27 (3.0e-27) & 1.3e-24 (4.4e-24) & 1.1e-21 (4.0e-21)\\
\hline
$\mathcal{O}_{8}$ [2 (1+0)] & 8.3e-30 (2.2e-29) & 9.0e-28 (2.3e-27) & 9.3e-25 (2.9e-24) & 3.4e-22 (1.2e-21)\\
\hline
$\mathcal{O}_{9}$ [2 (0+1)] & 5.1e-30 (1.3e-29) & 5.5e-28 (1.5e-27) & 6.8e-25 (2.3e-24) & 6.0e-22 (2.1e-21)\\
\hline
$\mathcal{O}_{10}$ [2 (0+1)] & 4.7e-30 (1.3e-29) & 6.0e-28 (1.6e-27) & 7.8e-25 (2.4e-24) & 6.1e-22 (2.2e-21)\\
\hline
$\mathcal{O}_{11}$ [2 (0+1)] & 4.1e-30 (1.1e-29) & 4.4e-28 (1.2e-27) & 3.4e-25 (1.1e-24) & 1.5e-23 (4.9e-23)\\
\hline
$\mathcal{O}_{12}$ [2 (1+0)] & 9.4e-30 (2.7e-29) & 1.1e-27 (3.1e-27) & 1.4e-24 (4.6e-24) & 1.2e-21 (4.0e-21)\\
\hline
$\mathcal{O}_{13}$ [4 (1+1)] & 2.2e-33 (6.2e-33) & 1.4e-29 (3.7e-29) & 6.3e-24 (1.9e-23) & 9.3e-21 (3.2e-20)\\
\hline
$\mathcal{O}_{14}$ [4 (1+1)] & 2.3e-33 (6.0e-33) & 1.4e-29 (3.6e-29) & 6.8e-24 (2.0e-23) & 1.0e-20 (3.4e-20)\\
\hline
$\mathcal{O}_{15}$ [6 (1+2)] &-- &-- & 2.5e-23 (7.9e-23) & 6.8e-20 (2.3e-19)\\
\hline
\end{tabular}
\caption{Upper limits on the scattering cross section $\sigma_p^{(i)}$ \eqref{eq:cs-simple} [evaluated at $v$$=$$(220~\km/\s)/c$] in units of $\cm^2$ at the 68\% (95\%) confidence level, as inferred from \textit{Planck} 2015 data. The DM spin is fixed to $S_\chi$=$1/2$. The first column indicates which operator is under study and lists its power-law dependence on the perpendicular component of velocity ($\alpha$) and the momentum transfer ($\beta$), as well as the power of relative velocity for the corresponding cross section, $n$=$2(\alpha+\beta)$.}
\label{tab:alllims}
\end{table}

We find no evidence of DM--proton scattering for $m_\chi$$>$$15~\keV$.
We thus report a complete set of cosmological upper limits on the coupling coefficients and cross sections associated with the 14 different operators in Eq.~\eqref{eq:operators}.
Our inferred upper limits on the cross sections for proton scattering, $\sigma_p^{(i)}$, with spin-$1/2$ DM are listed in Table~\ref{tab:alllims}.\footnote{The lowest-mass data points for $\mathcal{O}_{15}$ are missing. As we alluded to in a previous footnote, our code is not sufficiently optimized to produce accurate power spectra for the extremely strong velocity dependence of $n$=6 at low DM mass. We leave a detailed treatment of this regime for future work.}
These cross sections are evaluated at $v$=$(220~\km/\s)/c$, the relative velocity relevant for direct detection.
In the Appendix, Tables~\ref{tab:alllims-tilde} and \ref{tab:alllims-c2} list the upper limits in terms of the quantities $\widetilde\sigma_p^{(i)}$ and $c_i^2$, respectively, to allow for a more straightforward comparison to other works.
Note that since $\mathcal{O}_1$ is velocity independent, $\widetilde\sigma_p^{(1)}$$=$$\sigma_p^{(1)}$, consistent with the notation in our previous study~\cite{Gluscevic:2017ywp}.
The limits for DM spins other than $1/2$ are easily obtained by rescaling the cross section by $S_\chi (S_\chi + 1) / (3/4)$ (in accordance with Table~\ref{tab:Rresponse}) for any operator except $\mathcal{O}_{1}$, $\mathcal{O}_{3}$, $\mathcal{O}_{7}$, and $\mathcal{O}_{10}$, which are independent of DM spin.

\begin{figure*}[t]
\includegraphics[height=0.35\textwidth]{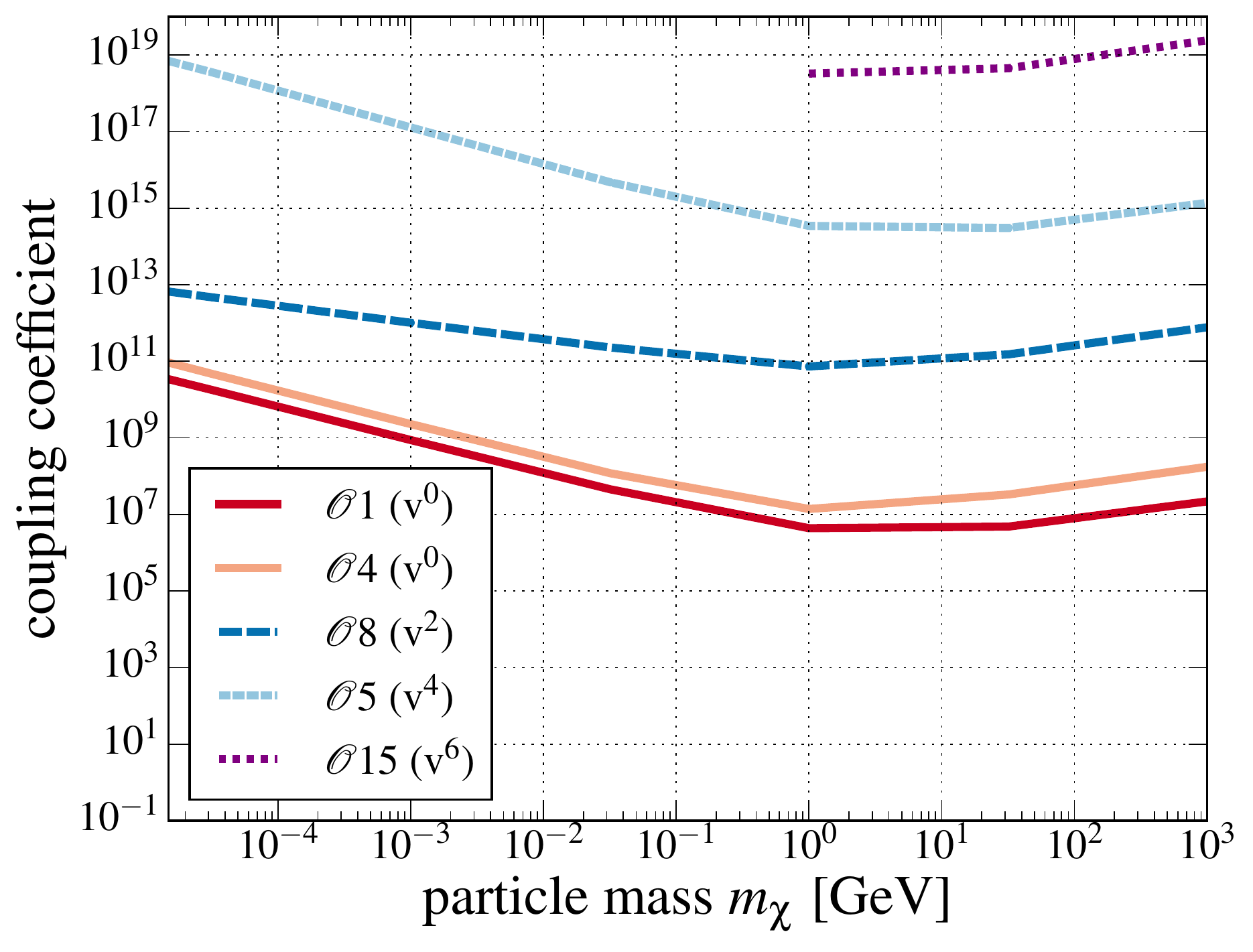}
\hspace{0.8cm}
\includegraphics[height=0.35\textwidth]{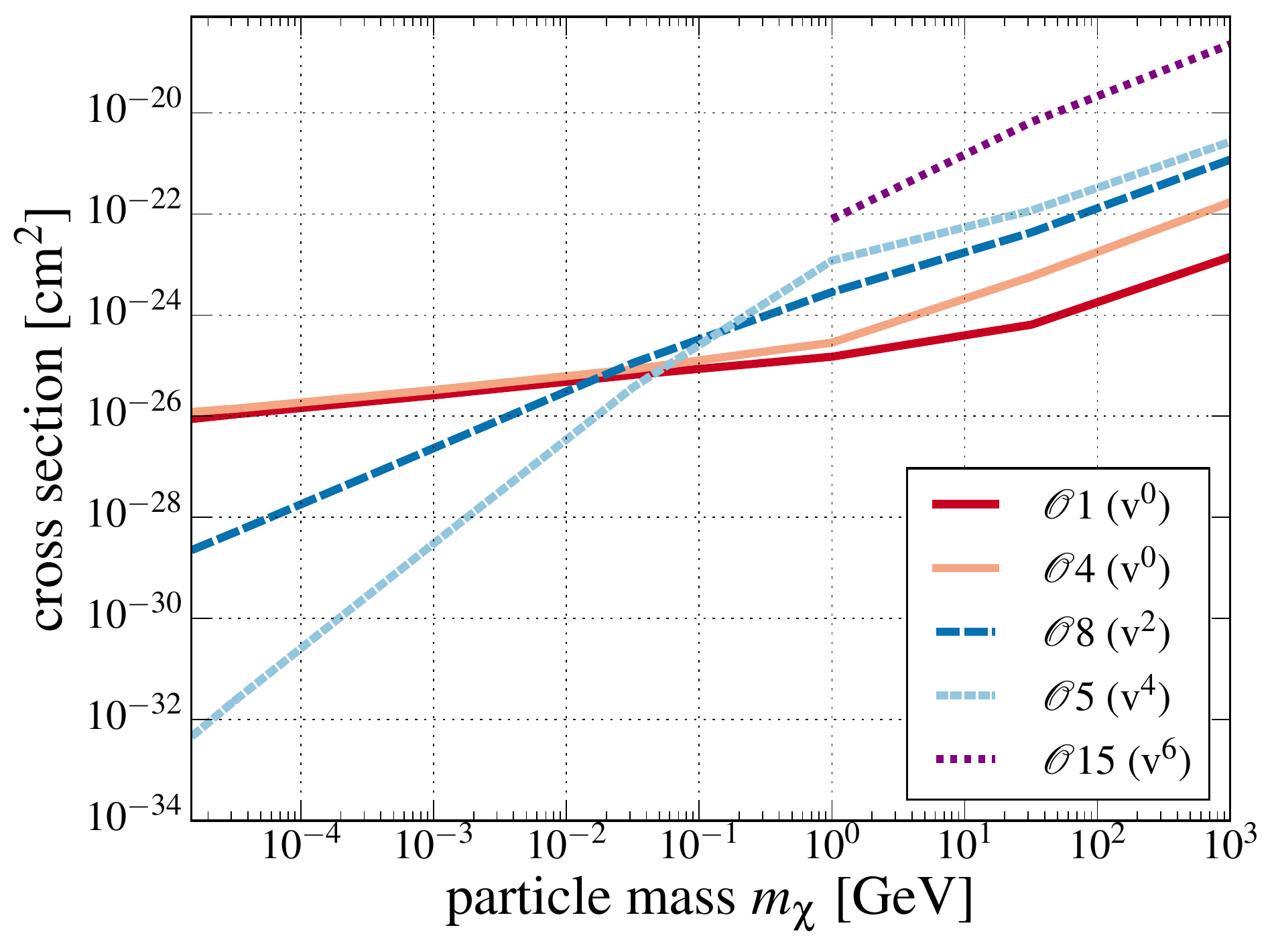}
\caption{The inferred upper limits on the DM--proton coupling coefficients and the corresponding cross sections [evaluated at $v$$=$$(220~\km/\s)/c$] for $\{\mathcal{O}_1, \mathcal{O}_5, \mathcal{O}_8, \mathcal{O}_{15}\}$, chosen to represent all classes of relative velocity scalings of the cross section (indicated in the legend). We also show limits for spin-dependent scattering through $\mathcal{O}_4$ to compare to spin-independent scattering through $\mathcal{O}_1$. Regions above the curves are excluded with the \textit{Planck} 2015 temperature, polarization, and lensing measurements at 95$\%$ confidence.\label{fig:exclusions}}
\end{figure*}

We choose the subset of operators $\{\mathcal{O}_1, \mathcal{O}_5, \mathcal{O}_8, \mathcal{O}_{15}\}$ to represent the various types of velocity-dependent interactions with $n$$\in$$\{0,2,4,6\}$, and we choose $\mathcal{O}_4$ as a representative of spin-dependent interactions.
For these operators, we sample the likelihood on a finer grid of logarithmically-spaced masses: $m_\chi$=$\{15~\keV$, $32~\keV$, $1~\MeV$, $32~\MeV$, $1~\GeV$, $32~\GeV$, $1~\TeV\}$.
We show the resulting 95$\%$ confidence-level exclusion curves in Figure~\ref{fig:exclusions}.
Finally, in Figures~\ref{fig:triangle1}, \ref{fig:triangle8}, \ref{fig:triangle5}, and \ref{fig:triangle15}, we show the posterior probability distributions of $\Lambda$CDM parameters and coupling coefficients for $\mathcal{O}_1$, $\mathcal{O}_8$, $\mathcal{O}_5$, and $\mathcal{O}_{15}$, respectively, for fixed $m_\chi$=$1~\GeV$ and $S_\chi$=$1/2$.

\section{Discussion and Conclusions}
\label{sec:conclusions}

We analyzed \textit{Planck} 2015 $TT$, $EE$, $TE$, and lensing power spectra in search for evidence of DM--proton interactions in the early Universe.
Our results are consistent with absence of interactions through operators dependent or independent of spin and velocity, and we thus report the first cosmological upper limits on the full nonrelativistic effective field theory of DM--proton scattering for masses above 15~\keV.
The main result is summarized in Figure~\ref{fig:exclusions}.
In this section, we discuss in more detail our results and their implications for future studies.

We first note that comparing the cross-section constraints for $\mathcal{O}_1$ and $\mathcal{O}_4$ (shown in the right panel of Figure~\ref{fig:exclusions}) illustrates the impact of including scattering with helium in our analyses.
These two operators represent the standard spin-independent and spin-dependent scattering with no dependence on relative velocity, and the main difference in their upper limits comes from the fact that DM cannot interact through $\mathcal{O}_4$ with helium, which has zero nuclear spin, while $\mathcal{O}_1$ interacts with both helium and hydrogen.
The coefficient $R_\chi$, defined in Eq.~\eqref{eq:rate}, quantifies the rate of momentum transfer between DM and all baryonic species that participate in a given interaction.
If there is only scattering with protons (as is the case for $\mathcal{O}_4$), the maximum of $R_\chi$ occurs for DM masses near a GeV, above which $R_\chi$ rapidly decreases (asymptoting to $\sim$$1/m_\chi$ behavior at $m_\chi$$\gg$1~\GeV), resulting in a loss of constraining power.
Including scattering on helium (for $\mathcal{O}_1$) shifts its maximum above a GeV (since helium is roughly four times as massive as the proton), thereby substantially improving the constraint at high mass.
As previously reported in Ref.~\cite{Gluscevic:2017ywp}, although helium contributes only about a quarter of the total baryonic mass, the modest shift of the maximum in $R_\chi$ amounts to an improvement of about a factor of 6 near $1~\TeV$ in sensitivity of CMB probes to spin-independent scattering, while it has no impact on constraints for spin-dependent scattering.

We now examine the constraints on the coupling coefficients in the left panel of Figure~\ref{fig:exclusions} for operators whose cross sections have different scalings of relative velocity.
There is a hierarchy of constraints: interactions with the stronger velocity dependencies are more suppressed in the regime of very low velocities (of interest to cosmological studies), so CMB observables have less constraining power on the coupling coefficients of such interactions.
It is also evident that CMB observations are most sensitive to DM masses around a $\GeV$, when considering interactions with protons.
Roughly speaking, the sensitivity drops at higher masses, because the amount of momentum transferred in the scattering process saturates ($\mu_{\chi B}$ approaches the mass of the proton) but the drag force per unit mass drops as $1/m_\chi$; the sensitivity drops at lower masses, because the momentum transfer scales as $m_\chi$ in that regime ($\mu_{\chi B}$ approaches $m_\chi$).
Understanding the detailed mass dependence in either panel of Figure~\ref{fig:exclusions} for a given operator is nontrivial and does not immediately follow from the analytic expressions of Secs.~\ref{sec:boltzmann} and \ref{sec:EFT}.
One reason is that the CMB observables are controlled by two quantities: the coefficients for the rate of momentum transfer $R_\chi$ and of heat transfer $R_\chi^\prime$.
The former appears in Eq.~\eqref{eq:fluct} and controls the impact DM interactions have on density and velocity fluctuations at a given redshift; the latter appears in Eq.~\eqref{eq:temp} and controls the time of thermal decoupling between the DM and baryon fluids.
Their scaling with $m_\chi$ is different, and the resulting mass dependence observed in Figure~\ref{fig:exclusions} follows from a combination of the two.

When comparing the left and right panels of Figure~\ref{fig:exclusions} or examining the entries of Table~\ref{tab:alllims}, it is important to keep in mind that we treat the coupling coefficients as the primary parameter to constrain from the data, for a fixed DM mass.
On the other hand, the scattering cross section is a derived quantity, obtained by the relation in Eq.~\eqref{eq:cs-simple}; it scales quadratically with the coupling coefficient, but its dependence on $m_\chi$ is determined by the operator at hand.
We emphasize that the low-mass limiting behavior of the exclusion curve for the cross section is not solely determined by $n$=$2(\alpha+\beta)$, the power of the cross section scaling with relative velocity.
From Eq.~\eqref{eq:cs-simple}, we see that velocity dependence in the cross section arises from the perpendicular component of the relative velocity (associated with the power $\alpha$) and the momentum transfer (associated with the power $\beta$) that appear in the DM response function; however, the momentum transfer is a function of $\mu_{\chi B}$, which introduces a mass scaling for the cross section that involves $\beta$ but not $\alpha$.
This dependence on $\beta$ explains the different low-mass behaviors of the exclusion curves associated with the coupling coefficients of different operators, shown in the left panel of Figure~\ref{fig:exclusions}.

In the high-mass limit of $m_\chi$$\gg$$1~\GeV$, the situation is greatly simplified.
The coefficient for the rate of heat transfer $R_\chi^\prime$ approaches that for the rate of momentum transfer $R_\chi$, and the thermal term $T_\chi / m_\chi$ for DM in Eq.~\eqref{eq:rate} is negligible, regardless of operator.
Thus, cosmological observables are controlled solely by $\sigma_B^{(i)} / m_\chi$.
Moreover, $\sigma_B^{(i)}$ in Eq.~\eqref{eq:cs-simple} scales as $c_i^2$ and has no $m_\chi$ dependence.
All exclusion curves in Figure~\ref{fig:exclusions} exhibit the behavior $\sigma_B^{(i)}$$\sim$$c_i^2$$\sim$$m_\chi$.
This relation allows extrapolation of our results in Table~\ref{tab:alllims} to arbitrarily high masses: the upper limit on the DM--proton cross section for $m_\chi$$>$$\TeV$ is obtained by scaling the upper limit at $1~\TeV$ by $m_\chi/\TeV$.

\begin{figure*}[hp]
\includegraphics[width=\textwidth]{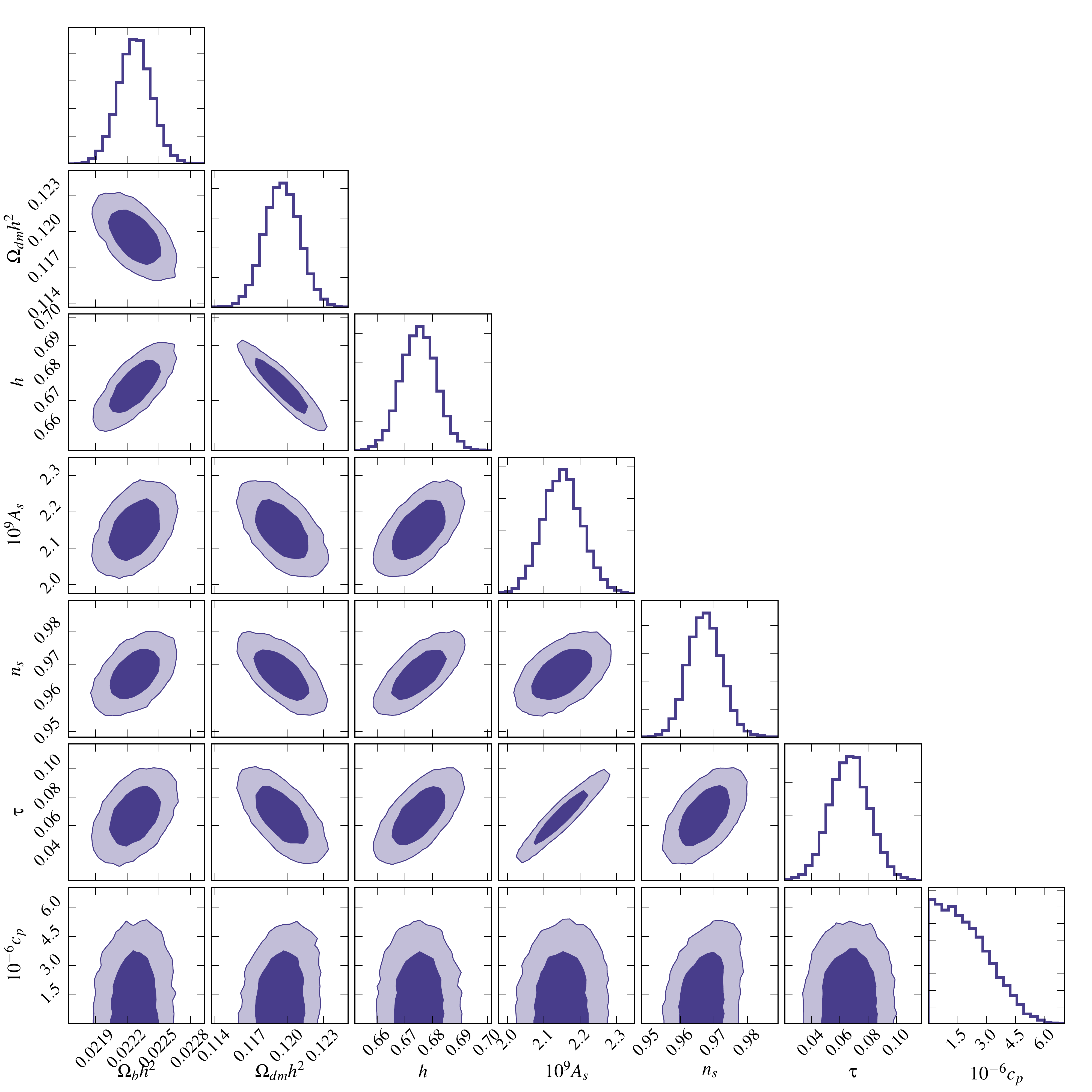}
\centering
\caption{The inferred posterior probability distribution for the $\Lambda$CDM parameters and the DM--proton coupling coefficient $c_p$=$c_1$ for velocity-independent scattering through $\mathcal{O}_1$, with DM mass $m_\chi$=$1$~\GeV and spin $S_\chi$=1/2. The posterior is shown at the 68$\%$ and 95$\%$ confidence levels, obtained from a joint analysis of \textit{Planck} 2015 temperature, polarization, and lensing power spectra. Marginalized posteriors are shown in the top panels of each column.\label{fig:triangle1}}
\end{figure*}

\begin{figure*}[hp]
\includegraphics[width=\textwidth]{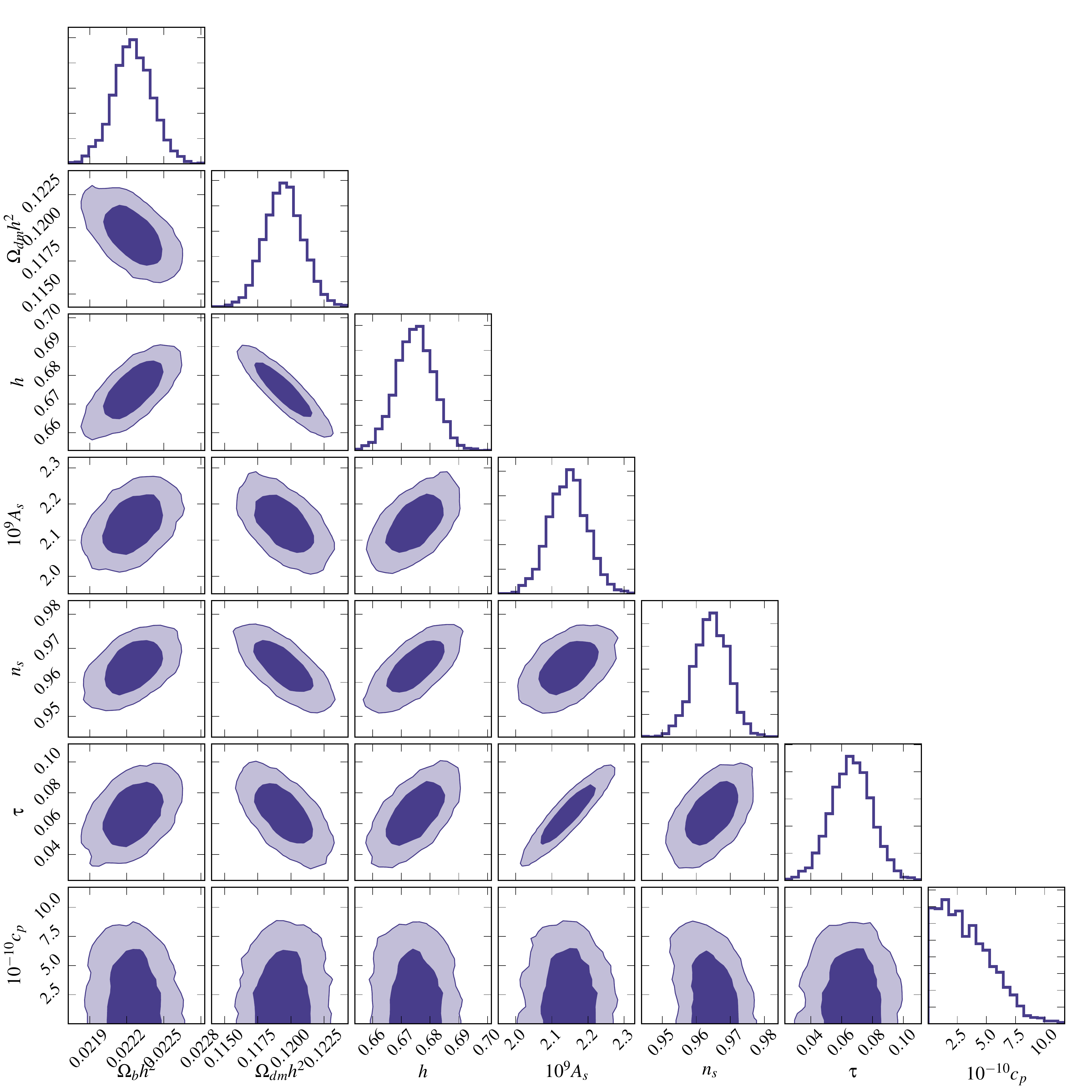}
\caption{The same as Figure~\ref{fig:triangle1}, except for $\mathcal{O}_8$.\label{fig:triangle8}}
\end{figure*}

\begin{figure*}[hp]
\includegraphics[width=\textwidth]{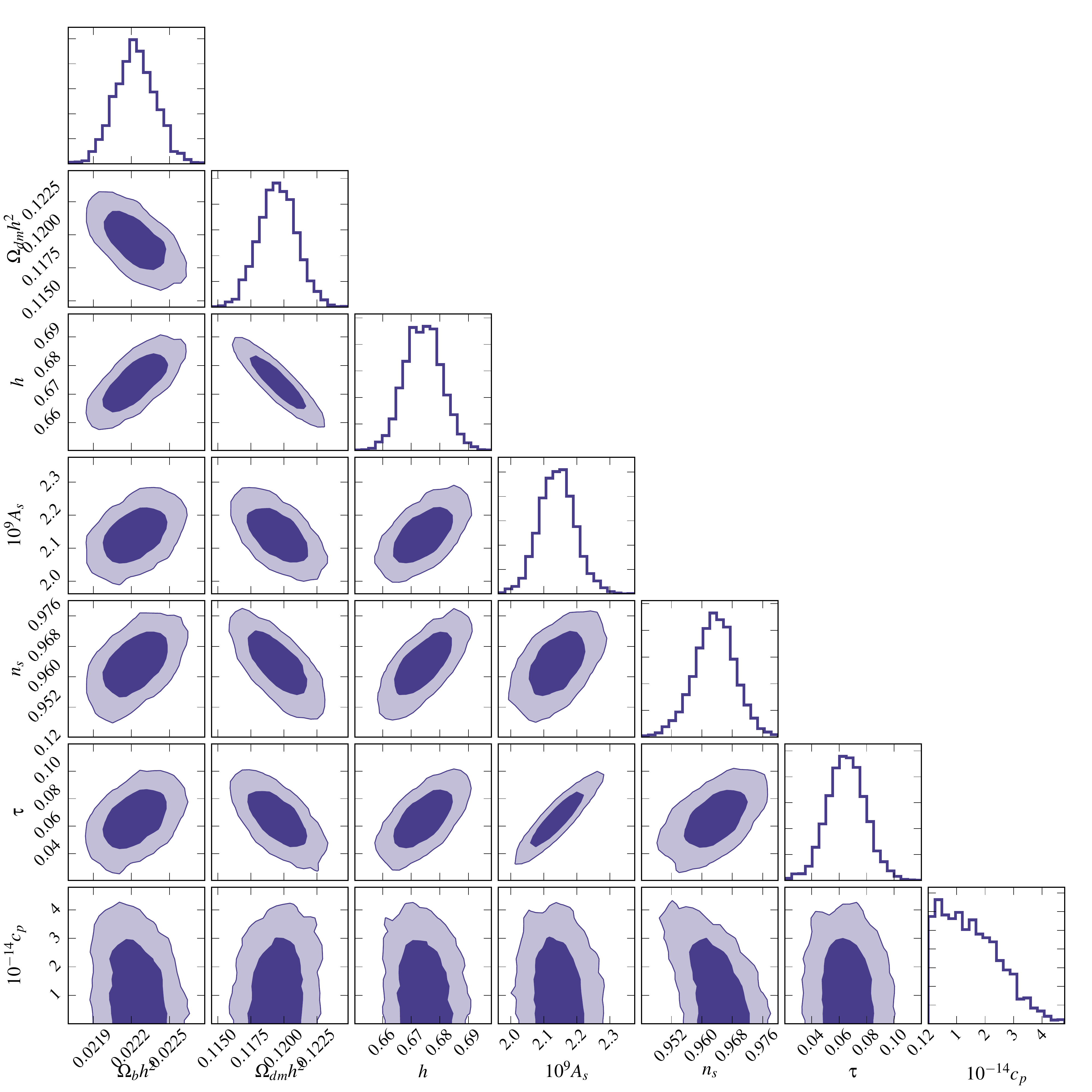}
\caption{The same as Figure~\ref{fig:triangle1}, except for $\mathcal{O}_5$.\label{fig:triangle5}}
\end{figure*}

\begin{figure*}[hp]
\includegraphics[width=\textwidth]{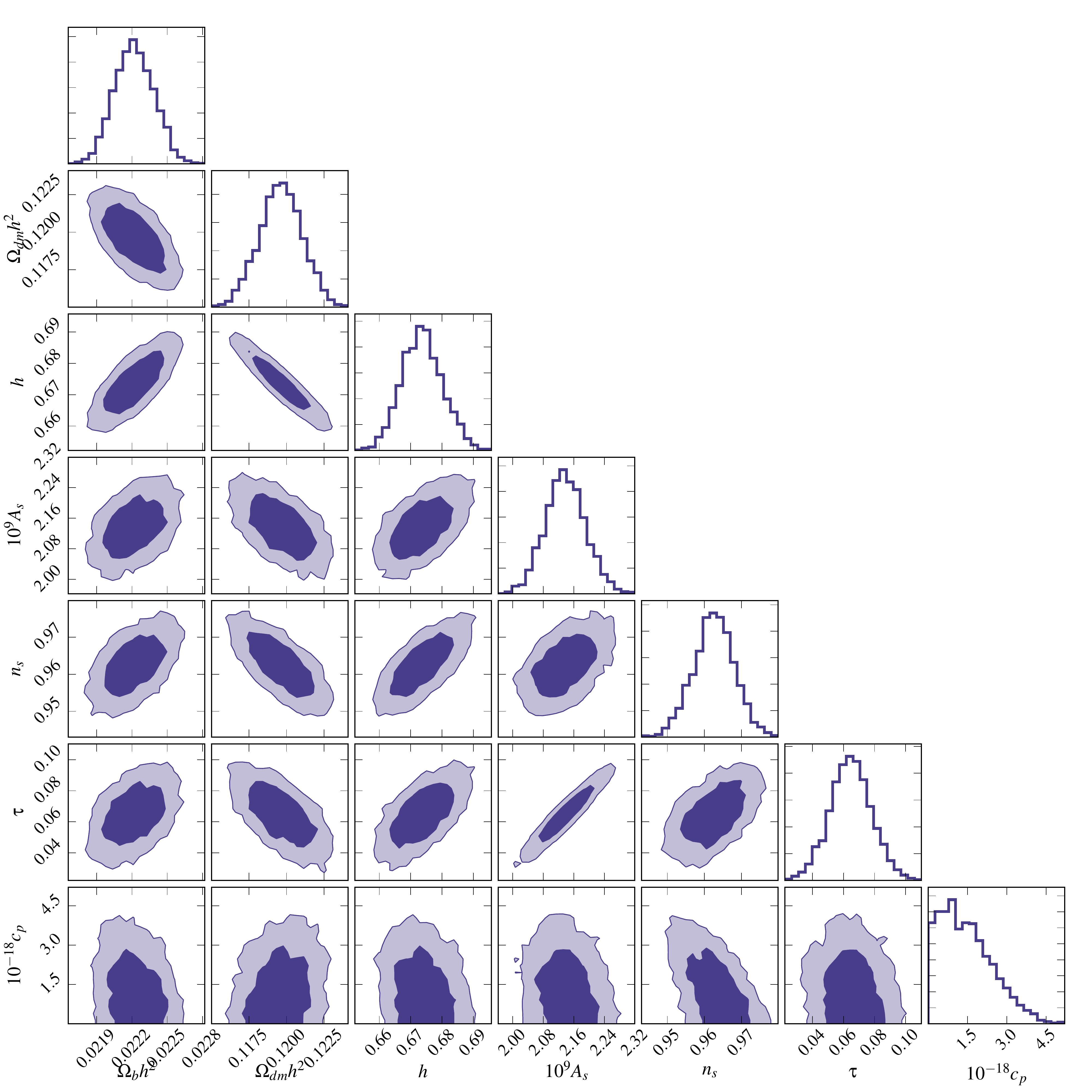}
\caption{The same as Figure~\ref{fig:triangle1}, except for $\mathcal{O}_{15}$.\label{fig:triangle15}}
\end{figure*}

Let us now take a look at the possible degeneracies between the DM scattering signal and other cosmological parameters.
The addition of high-$\ell$ polarization and lensing improves our reported constraint by about 30\%; thus, most of \textit{Planck}'s constraining power comes from the temperature anisotropy at the smallest angular scales measured with a high signal-to-noise ratio (where the \textit{Planck} error bars are their smallest in the left panel of Figure~\ref{fig:res}).
In this regime, where $TT$ measurements at $\ell$$\sim$1200 dominate the constraint, the effect of DM--proton scattering is not strongly degenerate with any of the standard cosmological parameters.
There is, however, a mild degeneracy with the scalar spectral index $n_s$, as seen in Figure~\ref{fig:triangle1} for $\mathcal{O}_1$.
The origin of this degeneracy can be understood from the left panel of Figure~\ref{fig:res}: increasing the strength of the DM--proton coupling progressively suppresses power on smaller scales, and a larger value of $n_s$ is needed to restore $C_\ell$ to within the measurement error, giving rise to the mild positive correlation between $c_1$ and $n_s$.
Analogous plots of the posterior probabilities for the cases of velocity-dependent scattering are shown in Figures~\ref{fig:triangle8}, \ref{fig:triangle5}, and \ref{fig:triangle15} for $\mathcal{O}_8$, $\mathcal{O}_5$, and $\mathcal{O}_{15}$, respectively.
The degeneracy with $n_s$ reverses sign when the leading effect of scattering on relevant angular scales ($\ell$$\sim$1200) gives rise to an enhancement of power rather than a suppression (such as the case for scattering through $\mathcal O_8$, for example, shown in Figure~\ref{fig:res}).

Finally, we briefly discuss future prospects for cosmological probes of DM interactions.
Given that the effect of DM--baryon interactions is stronger on smaller angular scales, high-resolution ground-based CMB observations from existing experiments (such as the Atacama Cosmology Telescope~\cite{2017JCAP...06..031L} and the South Pole Telescope~\cite{Simard:2017xtw}) and from future experiments (such as the Simons Observatory~\cite{Ade:2018sbj} and CMB-Stage 4~\cite{2016arXiv161002743A}) could improve upon our limits, particularly in the regime of sub-GeV DM.
As shown in the right panel of Figure~\ref{fig:res}, DM scattering is also imprinted on $P(k)$ and is progressively more prominent at larger values of $k$.
At the level of current constraints, DM scattering becomes inefficient at redshifts $z$$<$$10^4$ due to Hubble expansion, so its effect on $P(K)$ is similar to a $k$-dependent change in the initial conditions, for all effective operators considered in this work.
We thus expect that galaxy-survey and large-scale structure (LSS) measurements from BOSS~\cite{2013AJ....145...10D}, DES~\cite{2018arXiv180103181A}, LSST~\cite{2017arXiv170804058L}, DESI~\cite{2016arXiv161100036D}, etc.~can help improve constraints on DM interactions; see, for example, results from the Lyman-$\alpha$ data analysis in Ref.~\cite{Dvorkin:2013cea}.
Moreover, there is evidence that including the Lyman-$\alpha$ forest power spectrum leads to smaller values of the scalar spectral index (when also allowing the spectral index to run)~\cite{Palanque-Delabrouille:2015pga}, indicating that Lyman-$\alpha$ data may aid in constraining DM interactions on multiple fronts.
However, unlike the case of CMB, LSS analyses require modeling of complicated baryonic effects as well as treatments of nonlinearities and other systematic effects that arise on small scales; a calculation of the nonlinear effects depends on the cosmological context and has not yet been fully addressed within a cosmology with DM--baryon scattering, in which $P(k)$ exhibits oscillatory features shown in Figure~\ref{fig:res}.
For these reasons, the CMB is currently the most robust cosmological probe of DM--baryon interactions; we have thus only focused on CMB measurements and leave the extension to LSS analyses for future work.

In summary, this study paves a path for comprehensive and model-independent cosmological studies of low-energy DM physics.
Cosmological probes are particularly attuned to testing for the presence of sub-GeV dark matter, whose interactions may not be detectable in standard laboratory searches for weakly-interacting massive particles, but may leave an imprint in the early Universe.
Many forthcoming observations of the CMB and large-scale structure promise to further deepen the coverage of the relevant parameter space for DM candidates.

\afterpage{\clearpage}

\afterpage{\clearpage}

\afterpage{\clearpage}

\afterpage{\clearpage}

\newpage

\acknowledgments
VG gratefully acknowledges the support of the Eric Schmidt fellowship at the Institute for Advanced Study and the hospitality and support of Juna Kollmeier through the Visitor Program at Carnegie Observatories, where this work started.
The authors thank John Beacom, Joanna Dunkley, Marc Kamionkowski, and Timothy Morton for useful discussions.
Posterior probability distributions in this study were visualized using \texttt{corner.py}~\cite{corner}.

\section{Appendix}

In this Appendix, we present the 68\% and 95\% confidence-level upper limits on two quantities related to the scattering cross section $\sigma_p^{(i)}$ of Table~\ref{tab:alllims}, for a more straightforward comparison with other literature.
Table~\ref{tab:alllims-tilde} lists the upper limits on $\widetilde\sigma_p^{(i)}$, corresponding to quantity $\sigma_0$ in Ref.~\cite{Dvorkin:2013cea}, for example.
Table~\ref{tab:alllims-c2} lists the upper limits on $c_i^2$, which corresponds to the limits in Figure~6 of Ref.~\cite{Aprile:2017aas}, for example.

\begin{table}[h]
\begin{tabular}{ |c|c|c|c|c| }
\hline
\textbf{Operator} & \multicolumn{4}{|c|}{\textbf{DM Mass}} \\
\cline{2-5}
\textbf{[$n$ ($\alpha$+$\beta$)]} & \textbf{15 keV} & \textbf{1 MeV} & \textbf{1 GeV} & \textbf{1 TeV}\\
\hline
$\mathcal{O}_{1}$ [0 (0+0)] & 2.9e-27 (8.8e-27) & 9.1e-27 (2.6e-26) & 4.9e-26 (1.5e-25) & 4.7e-24 (1.4e-23)\\
\hline
$\mathcal{O}_{3}$ [4 (1+1)] & 7.8e-21 (2.0e-20) & 4.7e-17 (1.3e-16) & 2.2e-11 (6.6e-11) & 3.3e-08 (1.2e-07)\\
\hline
$\mathcal{O}_{4}$ [0 (0+0)] & 3.7e-27 (1.2e-26) & 1.1e-26 (3.3e-26) & 9.3e-26 (2.9e-25) & 5.6e-23 (1.7e-22)\\
\hline
$\mathcal{O}_{5}$ [4 (1+1)] & 6.4e-21 (1.6e-20) & 3.9e-17 (1.0e-16) & 1.4e-11 (4.1e-11) & 3.0e-09 (9.3e-09)\\
\hline
$\mathcal{O}_{6}$ [4 (0+2)] & 5.2e-21 (1.3e-20) & 3.3e-17 (8.4e-17) & 1.5e-11 (4.6e-11) & 2.2e-08 (7.3e-08)\\
\hline
$\mathcal{O}_{7}$ [2 (1+0)] & 1.9e-23 (5.1e-23) & 2.0e-21 (5.6e-21) & 2.5e-18 (8.3e-18) & 2.1e-15 (7.5e-15)\\
\hline
$\mathcal{O}_{8}$ [2 (1+0)] & 1.5e-23 (4.2e-23) & 1.7e-21 (4.3e-21) & 1.7e-18 (5.3e-18) & 6.4e-16 (2.2e-15)\\
\hline
$\mathcal{O}_{9}$ [2 (0+1)] & 9.4e-24 (2.5e-23) & 1.0e-21 (2.8e-21) & 1.3e-18 (4.3e-18) & 1.1e-15 (3.9e-15)\\
\hline
$\mathcal{O}_{10}$ [2 (0+1)] & 8.8e-24 (2.5e-23) & 1.1e-21 (2.9e-21) & 1.4e-18 (4.4e-18) & 1.1e-15 (4.1e-15)\\
\hline
$\mathcal{O}_{11}$ [2 (0+1)] & 7.5e-24 (2.0e-23) & 8.1e-22 (2.3e-21) & 6.3e-19 (2.0e-18) & 2.8e-17 (9.1e-17)\\
\hline
$\mathcal{O}_{12}$ [2 (1+0)] & 1.7e-23 (5.0e-23) & 2.1e-21 (5.7e-21) & 2.6e-18 (8.5e-18) & 2.1e-15 (7.5e-15)\\
\hline
$\mathcal{O}_{13}$ [4 (1+1)] & 7.7e-21 (2.1e-20) & 4.8e-17 (1.3e-16) & 2.2e-11 (6.6e-11) & 3.2e-08 (1.1e-07)\\
\hline
$\mathcal{O}_{14}$ [4 (1+1)] & 7.8e-21 (2.1e-20) & 4.9e-17 (1.3e-16) & 2.3e-11 (6.8e-11) & 3.5e-08 (1.2e-07)\\
\hline
$\mathcal{O}_{15}$ [6 (1+2)] &-- &-- & 1.6e-04 (5.0e-04) & 4.4e-01 (1.5e+00)\\
\hline
\end{tabular}
\caption{Upper limits on the scattering cross section $\widetilde{\sigma}_p^{(i)}$ \eqref{eq:cs-simple} in units of $\cm^2$ at the 68\% (95\%) confidence level, as inferred from \textit{Planck} 2015 data. The DM spin is fixed to $S_\chi$=$1/2$. The first column indicates which operator is under study and lists its power-law dependence on the perpendicular component of velocity ($\alpha$) and the momentum transfer ($\beta$), as well as the power of relative velocity for the corresponding cross section, $n$=$2(\alpha+\beta)$. To compare to previous CMB limits on the DM---proton interactions~\cite{Chen:2002yh,Sigurdson:2004zp,Dvorkin:2013cea}, the upper limits on the coefficient of the momentum-transfer cross section can be obtained by multiplying the limits reported here by $2(1+\beta)/(2+\alpha+\beta)$.}
\label{tab:alllims-tilde}
\end{table}
\begin{table}[h]
\begin{tabular}{ |c|c|c|c|c| }
\hline
\textbf{Operator} & \multicolumn{4}{|c|}{\textbf{DM Mass}} \\
\cline{2-5}
\textbf{[$n$ ($\alpha$+$\beta$)]} & \textbf{15 keV} & \textbf{1 MeV} & \textbf{1 GeV} & \textbf{1 TeV}\\
\hline
$\mathcal{O}_{1}$ [0 (0+0)] & 3.9e+20 (1.2e+21) & 2.7e+17 (7.7e+17) & 6.2e+12 (1.9e+13) & 1.6e+14 (4.8e+14)\\
\hline
$\mathcal{O}_{3}$ [4 (1+1)] & 2.4e+37 (6.1e+37) & 1.8e+33 (5.1e+33) & 6.3e+28 (1.9e+29) & 6.7e+30 (2.4e+31)\\
\hline
$\mathcal{O}_{4}$ [0 (0+0)] & 2.6e+21 (8.6e+21) & 1.7e+18 (5.3e+18) & 6.3e+13 (1.9e+14) & 1.0e+16 (3.1e+16)\\
\hline
$\mathcal{O}_{5}$ [4 (1+1)] & 2.0e+37 (4.9e+37) & 6.1e+33 (1.6e+34) & 3.9e+28 (1.2e+29) & 6.0e+29 (1.9e+30)\\
\hline
$\mathcal{O}_{6}$ [4 (0+2)] & 3.2e+46 (8.0e+46) & 2.3e+39 (5.8e+39) & 8.3e+28 (2.5e+29) & 2.2e+30 (7.3e+30)\\
\hline
$\mathcal{O}_{7}$ [2 (1+0)] & 2.0e+25 (5.4e+25) & 4.8e+23 (1.3e+24) & 2.5e+21 (8.4e+21) & 5.8e+23 (2.0e+24)\\
\hline
$\mathcal{O}_{8}$ [2 (1+0)] & 1.6e+25 (4.4e+25) & 4.0e+23 (1.0e+24) & 1.7e+21 (5.4e+21) & 1.7e+23 (6.0e+23)\\
\hline
$\mathcal{O}_{9}$ [2 (0+1)] & 1.9e+34 (5.2e+34) & 1.1e+29 (3.0e+29) & 2.4e+21 (8.2e+21) & 1.5e+23 (5.2e+23)\\
\hline
$\mathcal{O}_{10}$ [2 (0+1)] & 9.1e+33 (2.5e+34) & 5.8e+28 (1.5e+29) & 1.4e+21 (4.2e+21) & 7.6e+22 (2.8e+23)\\
\hline
$\mathcal{O}_{11}$ [2 (0+1)] & 7.8e+33 (2.1e+34) & 4.3e+28 (1.2e+29) & 6.0e+20 (1.9e+21) & 1.9e+21 (6.2e+21)\\
\hline
$\mathcal{O}_{12}$ [2 (1+0)] & 3.7e+25 (1.0e+26) & 1.0e+24 (2.7e+24) & 5.2e+21 (1.7e+22) & 1.2e+24 (4.0e+24)\\
\hline
$\mathcal{O}_{13}$ [4 (1+1)] & 9.6e+37 (2.6e+38) & 3.0e+34 (8.0e+34) & 2.5e+29 (7.5e+29) & 2.6e+31 (8.9e+31)\\
\hline
$\mathcal{O}_{14}$ [4 (1+1)] & 9.6e+37 (2.5e+38) & 3.1e+34 (7.9e+34) & 2.7e+29 (7.8e+29) & 2.8e+31 (9.5e+31)\\
\hline
$\mathcal{O}_{15}$ [6 (1+2)] &-- &-- & 3.4e+36 (1.1e+37) & 1.8e+38 (5.9e+38)\\
\hline
\end{tabular}
\caption{Upper limits on the square of the coupling coefficients $c_i^2$ (unitless) for DM-proton scattering, at the 68\% (95\%) confidence level, as inferred from \textit{Planck} 2015 data. The DM spin is fixed to $S_\chi$=$1/2$. The first column indicates which operator is under study and lists its power-law dependence on the perpendicular component of velocity ($\alpha$) and the momentum transfer ($\beta$), as well as the power of relative velocity for the corresponding cross section, $n$=$2(\alpha+\beta)$. Entries in this table correspond to those of Table~\ref{tab:alllims-tilde}.}
\label{tab:alllims-c2}
\end{table}

\bibliography{physics-refs}

\end{document}